\documentclass[preprint,pre,showpacs]{revtex4}
\usepackage{amscd}
\usepackage{amsfonts}
\usepackage{amsmath}
\usepackage{amssymb}
\usepackage{color}
\usepackage{graphicx}
\usepackage{graphics}
\usepackage{hyperref}
\usepackage{wrapfig}
\usepackage[T1]{fontenc}
\usepackage[latin1]{inputenc}
\def\>{\rangle}
\def\<{\langle}
\def\n{\nonumber}

\def\sc{\scriptsize}

\begin{document}
\title{The Clausius inequality beyond the weak coupling limit:\\The quantum Brownian
       oscillator revisited}
\author{Ilki Kim$^{1}$}
\email{hannibal.ikim@gmail.com}
\author{G\"{u}nter Mahler$^{2}$}
\affiliation{$^{1}$Department of Physics, North Carolina
A$\&$T State University, Greensboro, NC 27411\\
$^{2}$Institute of Theoretical Physics I, University of Stuttgart,
70550 Stuttgart, Germany}
\date{\today}
\begin{abstract}
We consider a quantum linear oscillator coupled at an arbitrary
strength to a bath at an arbitrary temperature. We find an exact
closed expression for the oscillator density operator. This state is
non-canonical but can be shown to be equivalent to that of an {\em
uncoupled} linear oscillator at an effective temperature
$T_{\text{\sc eff}}^{\star}$ with an effective mass and an effective
spring constant. We derive an effective Clausius inequality
$\delta{\mathcal Q}_{\text{\sc eff}}^{\star} \leq T_{\text{\sc
eff}}^{\star}\,dS$, where $\delta{\mathcal Q}_{\text{\sc
eff}}^{\star}$ is the heat exchanged between the effective (weakly
coupled) oscillator and the bath, and $S$ represents a thermal
entropy of the effective oscillator, being identical to the
von-Neumann entropy of the coupled oscillator. Using this inequality
(for a cyclic process in terms of a variation of the coupling
strength) we confirm the validity of the second law. For a fixed
coupling strength this inequality can also be tested for a process
in terms of a variation of either the oscillator mass or its spring
constant. Then it is never violated. The properly defined Clausius
inequality is thus more robust than assumed previously.
\end{abstract}
\maketitle
%
\section{Introduction}\label{sec:introduction}
Thermodynamics of small quantum objects coupled to quantum
environments in the low temperature regime has attracted
considerable interest as the need for a better theoretical
understanding increases in response to novel experimental
manipulation of such systems. In particular, the finite coupling
strength between system and environment gives rise to some quantum
subtleties and so can no longer be neglected (calling for methods
addressed by `quantum thermodynamics' \cite{SHE02,MAH04,CAP05})
whereas ordinary quantum statistical mechanics is intrinsically
based on a vanishingly small coupling between them.

At the heart of quantum thermodynamics, the foundational question as
to the validity of the second law of thermodynamics comes up. In
fact, with its challenge the applicability of thermodynamics is at
stake. So far, the validity of this basic law has extensively been
examined in the scheme of a quantum harmonic oscillator linearly
coupled to an independent-oscillator model of a heat bath (quantum
Brownian oscillator) in equilibrium at a low temperature $T$. It has
been argued here that there is a violation of the Clausius
inequality representing the second law \cite{ALL00,NIE02} in such a
way that $\delta{\mathcal Q} \not\leq T\,dS$ at $T \to 0$ with
respect to a variation of a Hamiltonian parameter of the coupled
oscillator, namely, either its mass or spring constant. In the above
relation, $\delta{\mathcal Q}$ is the heat and $dS$ is the entropy
change.

However, following the second law in its Kelvin-Planck form, which
states that it is impossible to devise a machine ({\em i.e.}, a heat
engine) which, operating in a cycle, produces no effect other than
the extraction of heat from a thermal energy reservoir and the
performance of an equal amount of work \cite{ANN02}, it has been
demonstrated that an apparent excess energy in the coupled
oscillator at zero temperature $(T=0)$ is less than the minimum
value of the work (equivalent to the Helmholtz free energy at a
constant temperature) to couple the free oscillator to a bath so
that the second law is not violated down to zero temperature
\cite{FOR06,KIM06}.

This result has been generalized to a cyclic process of coupling and
decoupling between the oscillator and a bath at an arbitrary
temperature by obtaining the positive-valuedness of the minimum work
needed for the coupling minus the maximum useful work obtainable
from the oscillator in the decoupling (the second law with respect
to a variation of the coupling strength) \cite{KIM07}. This
positive-valuedness is actually at its maximum at zero temperature
and asymptotically vanishes with increasing temperature, whereas the
classical counterpart would identically vanish at an arbitrary
temperature (even for a non-vanishing coupling). It was further
claimed here that this quantum behavior is associated with the
system-bath entanglement induced by the finite coupling strength
between them (clearly, the coupled total system ({\em i.e.}, the
coupled oscillator plus bath) is in a thermal state with (partial)
entanglement whereas the decoupled total system is simply in a
separable state). It has, indeed, been found that at zero
temperature the energy fluctuation in the coupled oscillator can
provide entanglement information \cite{BUE04,BUE05}. This claim was
supported by the numerical analysis of the system-bath negativity as
an exact entanglement measure \cite{LUT08} that the negativity
behavior versus temperature is in accordance with the above quantum
behavior of the second law up to the existence of the critical
temperature above which the negativity vanishes.

It has also been shown \cite{LUT08} that the Clausius inequality (in
terms of the equilibrium temperature of the {\em total} coupled
system and the von-Neumann entropy of the coupled oscillator) is
actually violated with respect to a variation of the mass of the
coupled oscillator (not a variation of the coupling strength); the
behavior of this violation versus the temperature is essentially
different from that of the system-bath negativity so that it has
been concluded that the system-bath entanglement is not responsible
for the violation of the Clausius inequality. However, as the
reduced equilibrium density operator of the coupled oscillator is
not in form of the canonical thermal state $\hat{\rho}_{\beta}
\propto e^{-\beta \hat{H}_s}$, there is not a well-defined {\em
local} temperature of the coupled oscillator (especially in the low
temperature limit) so that applying the equilibrium temperature of
the {\em total} coupled system for the violation of the Clausius
inequality for the subsystem (actually with respect to a variation
of a {\em local} parameter of the coupled oscillator, namely, either
its mass or spring constant) is not justified. Further, this
violation was actually based on the numerical findings \cite{LUT08}
that the heat $\delta{\mathcal Q}$ exchanged with a bath in a
reversible variation of the local parameter is always strictly
greater than $T\,dS$, which, however, does not satisfy the equality
condition of a well-defined Clausius inequality for the reversible
process.

On the other hand, introducing some generalized entropic measure and
using its maximum condition \cite{THU07} it has been shown that the
Clausius inequality obtained in some operational form is valid under
such a generalization \cite{ABE08}. However, this approach is not
directly applicable for the quantum Brownian oscillator since the
reduced density operator of the coupled oscillator in equilibrium
({\em cf}. equations (\ref{eq:united_expression1}) and
(\ref{eq:united_expression12})) is not in form of the stationary
state obtained from the maximum condition of this generalization.

In this paper we intend to resolve the above controversial issue by
introducing an effective Clausius inequality with no violation,
well-defined in the scheme of quantum Brownian oscillator. To do so,
we begin with considering the reduced density operator of the
coupled oscillator.

\section{Reduced density operator of the coupled oscillator}\label{sec:reduced_operator}
The quantum Brownian motion in consideration is described by the
model Hamiltonian
\begin{equation}\label{eq:total_hamiltonian1}
    \hat{H}\; =\; \hat{H}_s\, +\, \hat{H}_b\, +\, \hat{H}_{sb}\,,
\end{equation}
where
\begin{eqnarray}\label{eq:total_hamiltonian2}
    &\displaystyle \hat{H}_s\, =\, \frac{\hat{p}^2}{2 M} +
    \frac{k_0}{2}\,\hat{q}^2\;\;\; ;\;\;\;
    \hat{H}_b\, =\, \sum_{j=1}^N \left(\frac{\hat{p}_j^2}{2 m_j} +
    \frac{k_j}{2}\,\hat{x}_j^2\right)&\n\\
    &\displaystyle \hat{H}_{sb}\, =\, -\hat{q} \sum_{j=1}^N c_j\,\hat{x}_j\, +\, \hat{q}^2
    \sum_{j=1}^N \frac{c_j^2}{2\,k_j}\,,&
\end{eqnarray}
and the spring constants are $k_0 = M \omega_0^2$ and $k_j =
m_j\,\omega_j^2$. From the hermiticity of Hamiltonian, the coupling
constants $c_j$ are obviously real-valued. The total system is
assumed to be in the canonical thermal equilibrium state
$\hat{\rho}_{\beta} = e^{-\beta \hat{H}}/Z_{\beta}$ where $\beta =
1/(k_B\,T)$, and $Z_{\beta}$ is the partition function. From the
fluctuation-dissipation theorem \cite{CAL51,WEI99}, it is known
\cite{FOR85} that
\begin{eqnarray}
    \hspace*{-.5cm}\textstyle \<\hat{q}^2\>_{\beta} &=& \textstyle \frac{\hbar}{\pi}\,\int_0^{\infty} d\omega\,
    \coth\left(\frac{\beta \hbar \omega}{2}\right)\,
    \text{Im}\{\tilde{\chi}(\omega +
    i\,0^+)\}\label{eq:x_correlation1}\\
    \hspace*{-.5cm}\textstyle \<\hat{p}^2\>_{\beta} &=& \textstyle \frac{M^2 \hbar}{\pi}\,\int_0^{\infty} d\omega\,
    \omega^2\,\coth\left(\frac{\beta \hbar \omega}{2}\right)\,
    \text{Im}\{\tilde{\chi}(\omega + i\,0^+)\}\label{eq:x_dot_correlation1}
\end{eqnarray}
in terms of the susceptibility
\begin{equation}\label{eq:susceptibility2}
    \tilde{\chi}(\omega)\; =\; -\frac{1}{M}\, \frac{\displaystyle
    \prod_{j=1}^N\, (\omega^2 - \omega_j^2)}{\displaystyle \prod_{k=0}^N\, (\omega^2 -
    \bar{\omega}_k^2)}
\end{equation}
where $\{\bar{\omega}_k\}$ are the normal-mode frequencies of the
total system $\hat{H}$. For an uncoupled oscillator,
$\text{Im}\,\tilde{\chi}(\omega+i0^+)$ obviously reduces to
$\frac{\pi}{2 M \omega_0} \delta(\omega-\omega_0)$ and thus
$\<\hat{q}^2\>_{\beta} = \frac{\hbar}{2 M \omega_0} \coth\frac{\beta
\hbar \omega_0}{2}$ and $\<p^2\>_{\beta} = \frac{M \hbar
\omega_0}{2} \coth\frac{\beta \hbar \omega_0}{2}$. For the
well-known Drude model (with a cut-off frequency $\omega_d$ and a
damping parameter $\gamma_o$), which is a prototype for physically
realistic damping, we have \cite{KIM07}
\begin{eqnarray}
    \<\hat{q}^2\>_{\beta}^{(d)} &=& {\textstyle \frac{1}{M}}
    \sum_{l=1}^3 {\textstyle \lambda_d^{(l)}\,\left\{\frac{1}{\beta \underline{\omega_l}}\,
    +\, \frac{\hbar}{\pi}\; \psi\left(\frac{\beta \hbar \underline{\omega_l}}{2 \pi}\right)\right\}}\label{eq:x_drude}\\
    \<\hat{p}^2\>_{\beta}^{(d)} &=& {\textstyle -M}
    \sum_{l=1}^3 {\textstyle \lambda_d^{(l)}\,\underline{\omega_l}^2\,\left\{\frac{1}{\beta \underline{\omega_l}}\,
    +\, \frac{\hbar}{\pi}\; \psi\left(\frac{\beta \hbar \underline{\omega_l}}{2 \pi}\right)\right\}}\label{eq:p_drude}
\end{eqnarray}
in terms of the digamma function $\psi(y) = d\,\ln\Gamma(y)/dy$
\cite{ABS74}, where $\underline{\omega_1} = \Omega$,
$\underline{\omega_2} = z_1$, $\underline{\omega_3} = z_2$, and the
coefficients
\begin{equation}\label{eq:coefficients}
    \textstyle\lambda_d^{(1)}\; =\; \frac{z_1\,+\,z_2}{(\Omega\,-\,z_1) (z_2\,-\,\Omega)}\; ;\;
    \lambda_d^{(2)}\; =\; \frac{\Omega\,+\,z_2}{(z_1\,-\,\Omega)
    (z_2\,-\,z_1)}\; ;\;
    \lambda_d^{(3)}\; =\; \frac{\Omega\,+\,z_1}{(z_2\,-\,\Omega) (z_1\,-\,z_2)}\,.
\end{equation}
Here we have adopted, in place of $(\omega_0, \omega_d, \gamma_o)$,
the parameters $({\mathbf w}_0, \Omega, \gamma)$ through the
relations \cite{FOR06}
\begin{equation}\label{eq:parameter_change0}
    \textstyle\omega_0^2\, :=\, {\mathbf w}_0^2\; \frac{\Omega}{\Omega\, +\, \gamma}\;\; ;\;\;
    \omega_d\, :=\, \Omega\, +\, \gamma\;\; ;\;\;
    \gamma_o\, :=\, \gamma\, \frac{\Omega\, (\Omega\, +\, \gamma)\,
    +\, {\mathbf w}_0^2}{(\Omega\, +\, \gamma)^2}\,,
\end{equation}
and then $z_1 = \gamma/2 + i {\mathbf w}_1$ and $z_2 = \gamma/2 - i
{\mathbf w}_1$ with ${\mathbf w}_1 = \sqrt{({\mathbf w}_0)^2 -
(\gamma/2)^2}$.

The equilibrium density operator of the coupled oscillator is known
as \cite{WEI99,ING88}
\begin{equation}\label{eq:density_operator1}
    \textstyle \<q|\hat{\rho}_s|q'\>\, =\, \frac{1}{\sqrt{2\pi \<\hat{q}^2\>_{\beta}}}\, \exp\left(-\frac{(q + q')^2}{8\,\<\hat{q}^2\>_{\beta}} -
    \frac{\<\hat{p}^2\>_{\beta}\,(q - q')^2}{2 \hbar^2}\right)\,.
\end{equation}
For an uncoupled oscillator this easily reduces to a well-known
expression \cite{FEY98}
\begin{eqnarray}\label{eq:density_operator2}
    \hspace*{-.5cm}\textstyle \<q|\hat{\rho}_{\beta}|q'\> &=& \textstyle \sqrt{\frac{c^2}{\pi}\,\tanh\frac{\beta \hbar
    \omega_0}{2}}\, \exp\left(-\frac{c^2}{4}\left\{(q + q')^2
    \tanh\left(\frac{\beta \hbar \omega_0}{2}\right) + (q - q')^2 \coth\left(\frac{\beta \hbar
    \omega_0}{2}\right)\right\}\right)
\end{eqnarray}
where the parameter
\begin{equation}\label{eq:parameter_c}
    \textstyle c\, =\, \sqrt{\frac{M \omega_0}{\hbar}}\,.
\end{equation}
Let us now derive a closed form of the matrix elements
\begin{equation}\label{eq:wahrscheinlichkeit1}
    \textstyle \rho_{nm}\, :=\, \textstyle \<n|\hat{\rho}_s|m\>\,
    =\, \textstyle \int_{-\infty}^{\infty}\int_{-\infty}^{\infty} dq\, dq'\,
    \psi_n^{\ast}(q)\, \<q|\hat{\rho}_s|q'\>\, \psi_m(q')
\end{equation}
in the basis composed of the eigenstates $\{|n\>, |m\>\}$ of an
uncoupled oscillator to confirm its deviation from a (diagonal) form
of the canonical thermal state $\hat{\rho}_{\beta}$. After making
lengthy calculations, every single step of which is provided in a
detail in Appendix \ref{sec:appendix1}, we arrive at the closed
expressions
\begin{subequations}
\begin{eqnarray}
    \rho_{2k,2{\mathit l}+1} &=& \rho_{2k+1,2{\mathit l}}\, =\, 0\label{eq:prob01_zero}\\
    \rho_{2k,2{\mathit l}} &=& \textstyle\frac{\left(-\Upsilon_{\beta}\right)^{k+{\mathit l}}}{c\;\sqrt{2\pi\,\<\hat{q}^2\>_{\beta}\,A_{\beta}}}\,
    \sqrt{\frac{\Gamma\left(k+\frac{1}{2}\right)}{k!} \frac{\Gamma\left({\mathit l}+\frac{1}{2}\right)}{{\mathit l}!}}\, {}_{2}\hspace*{-.05cm}F_1\left(-k,-{\mathit l}; \frac{1}{2};
    \frac{1}{\Delta_{\beta}}\right)\label{eq:prob1}\\
    \rho_{2k+1,2{\mathit l}+1} &=& \textstyle\frac{2\,\Lambda_{\beta}\,\left(-\Upsilon_{\beta}\right)^{k+{\mathit l}}}{c\;\sqrt{2\pi\,\<\hat{q}^2\>_{\beta}\,A_{\beta}}}\,
    \sqrt{\frac{\Gamma\left(k+\frac{3}{2}\right)}{k!} \frac{\Gamma\left({\mathit l}+\frac{3}{2}\right)}{{\mathit l}!}}\, {}_{2}\hspace*{-.05cm}F_1\left(-k,-{\mathit l}; \frac{3}{2};
    \frac{1}{\Delta_{\beta}}\right)\label{eq:prob11}
\end{eqnarray}
\end{subequations}
in terms of the hypergeometric function
${}_{2}\hspace*{-.05cm}F_1(a,b; c; z)$ \cite{ABS74}. Here, the four
dimensionless quantities
\begin{equation}\label{eq:prob2}
    \textstyle A_{\beta}\, =\, \frac{\left(c^2 + \frac{2 \<\hat{p}^2\>_{\beta}}{\hbar^2}\right)\left(c^2 + \frac{1}{2 \<\hat{q}^2\>_{\beta}}\right)}{4 c^4}\;
    ;\; \textstyle \Upsilon_{\beta}\, =\, \frac{\frac{\<\hat{p}^2\>_{\beta}}{\hbar^2 \<\hat{q}^2\>_{\beta}}\,-\,c^4}{4 c^4 A_{\beta}}\;
    ;\; \textstyle \Lambda_{\beta}\, =\, \frac{\frac{\<\hat{p}^2\>_{\beta}}{\hbar^2} - \frac{1}{4 \<\hat{q}^2\>_{\beta}}}{2 c^2 A_{\beta}}\;
    ;\; \textstyle \Delta_{\beta}\, =\, \left(\frac{\Upsilon_{\beta}}{\Lambda_{\beta}}\right)^2\,.
\end{equation}
As shown, the reduced density matrix
$\left(\hat{\rho}_s\right)_{nm}$ is symmetric with respect to
$(n,m)$. Subsequently we also find ({\em cf}. Appendix
\ref{sec:appendix1}) that for $k \geq l$
\begin{subequations}
\begin{eqnarray}
    \hspace*{-.5cm}\rho_{2k,2{\mathit l}} &=& \textstyle\frac{\left(-\Upsilon_{\beta}\right)^{k-{\mathit l}}\,\left(\Lambda_{\beta}\right)^{2{\mathit l}}\,\left(1-\Delta_{\beta}\right)^{\mathit l}}{c\;\sqrt{2\,\<\hat{q}^2\>_{\beta}\,A_{\beta}}}\,
    \sqrt{\frac{\Gamma\left(k+\frac{1}{2}\right)}{\Gamma\left({\mathit l}+\frac{1}{2}\right)} \frac{k!}{{\mathit l}!}}\, \frac{\Gamma(2{\mathit l}+1)}{\Gamma(k+{\mathit
    l}+1)}\;
    P_{2{\mathit l}}^{(k-{\mathit l}, k-{\mathit l})}\left(\frac{1}{\sqrt{1-\Delta_{\beta}}}\right)\label{eq:prob20}\\
    \hspace*{-.5cm}\rho_{2k+1,2{\mathit l}+1} &=& \textstyle\frac{\left(-\Upsilon_{\beta}\right)^{k-{\mathit l}}\,\left(\Lambda_{\beta}\right)^{2{\mathit l}+1}\,\left(1-\Delta_{\beta}\right)^{{\mathit l}+\frac{1}{2}}}{c\;\sqrt{2\,\<\hat{q}^2\>_{\beta}\,A_{\beta}}}\,
    \sqrt{\frac{\Gamma\left(k+\frac{3}{2}\right)}{\Gamma\left({\mathit l}+\frac{3}{2}\right)} \frac{k!}{{\mathit l}!}}\, \frac{\Gamma(2{\mathit l}+2)}{\Gamma(k+{\mathit
    l}+2)}\;
    P_{2{\mathit l}+1}^{(k-{\mathit l}, k-{\mathit l})}\left(\frac{1}{\sqrt{1-\Delta_{\beta}}}\right)\label{eq:prob21}
\end{eqnarray}
\end{subequations}
in terms of the Jacobi polynomial \cite{ABS74} and for $k < {\mathit
l}$ the matrix elements $\rho_{2k,2{\mathit l}}$ and
$\rho_{2k+1,2{\mathit l}+1}$ correspond, respectively, to
(\ref{eq:prob20}) and (\ref{eq:prob21}) with exchange of $k$ and
${\mathit l}$. These are easily united into such a single expression
that for either $(n \text{ even} \geq m \text{ even})$ or $(n \text{
odd} \geq m \text{ odd})$,
\begin{equation}\label{eq:united_expression1}
    \textstyle\left(\hat{\rho}_s\right)_{nm}\, =\,
    \frac{\left(-\Upsilon_{\beta}\right)^{\frac{n-m}{2}}\,\left(\Lambda_{\beta}\right)^{m+\frac{1}{2}}\,\left(\sqrt{1-\Delta_{\beta}}\right)^m}{\sqrt{\frac{\<\hat{q}^2\>_{\beta}\,\<\hat{p}^2\>_{\beta}}{\hbar^2}\,-\,\frac{1}{4}}}\,\sqrt{\frac{\Gamma\left(\left[\frac{n+1}{2}\right]+\frac{1}{2}\right)}{\Gamma\left(\left[\frac{m+1}{2}\right]+\frac{1}{2}\right)} \frac{\left[\frac{n}{2}\right]!}{\left[\frac{m}{2}\right]!}}\,
    \frac{\Gamma(m+1)}{\Gamma\left(\frac{n+m}{2}+1\right)}\;
    P_m^{\left(\frac{n-m}{2},\frac{n-m}{2}\right)}\left(\frac{1}{\sqrt{1-\Delta_{\beta}}}\right)\,,
\end{equation}
and for either $(n \text{ even} < m \text{ even})$ or $(n \text{
odd} < m \text{ odd})$ the matrix elements
$\left(\hat{\rho}_s\right)_{nm}$ are obviously given by
(\ref{eq:united_expression1}) with exchange of $m$ and $n$, where
$[y]$ represents the greatest integer less than or equal to $y$. On
the other hand, from (\ref{eq:prob01_zero}),
\begin{equation}\label{eq:united_expression12}
    \textstyle\left(\hat{\rho}_s\right)_{nm}\, =\, 0
\end{equation}
where either $(n \text{ even}, m \text{ odd})$ or $(n \text{ odd}, m
\text{ even})$. As seen, the reduced density matrix
$(\hat{\rho}_s)_{nm}$ is, in general, not in diagonal form of the
canonical thermal state $e^{-\beta \hbar \omega_0
(n+\frac{1}{2})}/Z_{\beta}$ being valid for an uncoupled oscillator,
where $\Lambda_{\beta} \to e^{-\beta \hbar \omega_0}$ and
$\Delta_{\beta} \to 0$. This confirms that there is not a
well-defined {\em local} temperature when one keeps starring at the
oscillator $\hat{H}_s$, hence ignoring that it (strongly) couples to
a bath ({\em cf}. Section \ref{sec:eigenvalue}, in which, on the
other hand, a well-defined effective local temperature is
introduced). We note here, however, that by using $\hat{q} =
\sqrt{\frac{\hbar}{2 m \omega_0}}\,(\hat{a} + \hat{a}^{\dagger})$,
$\hat{p} = i \sqrt{\frac{m \hbar \omega_0}{2}}\,(\hat{a}^{\dagger} -
\hat{a})$, and the matrix elements (\ref{eq:united_expression12})
with $\hat{a} |n\> = \sqrt{n}\,|n-1\>$ and $\hat{a}^{\dagger} |n\> =
\sqrt{n+1}\,|n+1\>$, as is the case for an uncoupled oscillator,
\begin{equation}\label{eq:expectation_value1}
    {\textstyle\<\hat{q}\>_{\beta}\, =\, \mbox{Tr}_s(\hat{q}\,\hat{\rho}_s)\,
    =\, \sqrt{\frac{\hbar}{2m \omega_0}}}\,\sum_{n=0}^{\infty}
    \textstyle\left(\sqrt{n}\,\rho_{n,n-1} + \sqrt{n+1}\,\rho_{n,n+1}\right)\, =\, 0
\end{equation}
and likewise $\<\hat{p}\>_{\beta} = 0$. Actually, it can
straightforwardly be verified that $\<\hat{q}^l\>_{\beta} =
\<\hat{p}^l\>_{\beta} = 0$ with $l$ odd.

Let us consider the probability of finding the $n$th eigenstate from
the coupled oscillator in $\hat{\rho}_s$, which reads
\begin{equation}\label{eq:probability1}
    \textstyle p_n\, =\, \left(\hat{\rho}_s\right)_{nn}\, =\,
    \frac{\left(\Lambda_{\beta}\right)^{n+\frac{1}{2}}\,\left(\sqrt{1-\Delta_{\beta}}\right)^n}{\sqrt{\frac{\<\hat{q}^2\>_{\beta}\,\<\hat{p}^2\>_{\beta}}{\hbar^2}\,-\,\frac{1}{4}}}\;
    P_n\left(\frac{1}{\sqrt{1-\Delta_{\beta}}}\right)
\end{equation}
in terms of the Legendre polynomial \cite{ABS74}
\begin{equation}\label{eq:legendre_polynomial1}
    P_n(z)\, =\, P_{n}^{(0,0)}(z)\, =\, \frac{1}{2^n} {\displaystyle
    \sum_{k=0}^{[\frac{n}{2}]}}\textstyle
    (-1)^k\,\binom{n}{k}\,\binom{2n-2k}{n}\,z^{n-2k}\,.
\end{equation}
Here the normalization $\sum_n p_n = 1$ easily appears with the aid
of (\ref{eq:prob2}) and the relation \cite{ABS74}
\begin{equation}\label{eq:probability3}
    \sum_{n=0}^{\infty}\textstyle P_n(z)\,h^n\; =\; \frac{1}{\sqrt{1 - 2 z h + h^2}}
\end{equation}
where $z = \frac{1}{\sqrt{1-\Delta_{\beta}}}$ and $h =
\Lambda_{\beta} \sqrt{1-\Delta_{\beta}}$. Then the internal energy
of the coupled oscillator is
\begin{equation}\label{eq:hamiltonian_internal_energy1}
    U_s\, =\, \<\hat{H}_s\>_{\beta}\, =\, \sum_n p_n E_n\, =\,\textstyle \frac{\<\hat{p}^2\>_{\beta}}{2
    M}\,+\,\frac{M \omega_0^2}{2}\,\<\hat{q}^2\>_{\beta}\,.
\end{equation}
A comment deserves here. In \cite{BUE04} and then \cite{BUE08}, each
of diagonal elements $W_n$ was obtained in terms of the Legendre
polynomial $P_n$ like in (\ref{eq:probability1}). On the other hand,
the closed expression for off-diagonal elements
$\left(\hat{\rho}_s\right)_{nm}$ in (\ref{eq:united_expression1})
and (\ref{eq:united_expression12}) has not been known, while by
means of the numerical integration of (\ref{eq:wahrscheinlichkeit1})
the off-diagonal elements $|\left(\hat{\rho}_s\right)_{nm}|$ have
been obtained for $m, n \leq 10$ in \cite{BUE08}. We will next
consider the eigenvalues and eigenstates of the reduced density
operator $\hat{\rho}_s$.

\section{Eigenvalue problem for the oscillator density operator}\label{sec:eigenvalue}
The eigenvalue problem to be solved reads
\begin{equation}\label{eq:eigenvalue_problem1}
    \textstyle\text{EV}_n\, :=\, \int_{-\infty}^{\infty} dq'\, \<q|\hat{\rho}_s|q'\>\,
    \phi_n(q')\, \stackrel{!}{=}\, p_n\, \phi_n(q)
\end{equation}
where the matrix element $\<q|\hat{\rho}_s|q'\>$ is given in
(\ref{eq:density_operator1}), and the eigenvalue $p_n$ is the
probability of finding the $n$th eigenstate of the {\em coupled}
oscillator. Following the idea used in \cite{SRE93}, we put an
ansatz
\begin{equation}\label{eq:guess_eigen_function1}
    \textstyle\phi_n(q')\, =\, \sqrt{\frac{\tilde{c}}{2^n n! \sqrt{\pi}}}\, e^{-\frac{\tilde{c}^2}{2} q^2}\, H_n(\tilde{c}\,q')
\end{equation}
({\em cf}. (\ref{eq:eigenfunction1})) into the integral in
(\ref{eq:eigenvalue_problem1}), which will yield
\begin{equation}\label{eq:eigenvalue_problem2}
    \textstyle\text{EV}_n\, =\, \sqrt{\frac{\tilde{c}}{2^n n!
    \sqrt{\pi}}}\, \frac{1}{\sqrt{\pi}\,\tilde{v}}\,
    \exp\left(-\left\{\frac{\tilde{c}^2}{2} + \left(\frac{\<\hat{p}^2\>_{\beta}}{\hbar^2\,\<\hat{q}^2\>_{\beta}} - \tilde{c}^4\right)
    \frac{\<\hat{q}^2\>_{\beta}}{2\,\tilde{v}^2}\right\} q^2\right) \times (\text{I})\,.
\end{equation}
Here $\tilde{v} = \sqrt{\frac{1}{4} + \tilde{c}^2
\<\hat{q}^2\>_{\beta} + v^2}$ with $v = \frac{1}{\hbar}
\sqrt{\<\hat{q}^2\>_{\beta}\,\<\hat{p}^2\>_{\beta}}$, and
\begin{equation}\label{eq:integral_hermite1}
    \textstyle (\text{I})\, :=\, \int_{-\infty}^{\infty}
    d\tilde{q}'\, e^{-(\tilde{q}' - y)^2}\, H_n(s\,\tilde{q}')
\end{equation}
where the dimensionless quantities
\begin{equation}\label{eq:eigenvalue_problem3}
    \textstyle \tilde{q}'\, =\, \frac{\tilde{v}}{\sqrt{2\,\<\hat{q}^2\>_{\beta}}}\, q'\; ;\;
    y\, =\, \frac{\sqrt{\<\hat{p}^2\>_{\beta}}}{\sqrt{2}\,\hbar\,\tilde{v}}
    \left(v - \frac{1}{4 v}\right)\, q\; ;\;
    s\, =\, \frac{\tilde{c}}{\tilde{v}} \sqrt{2\,\<\hat{q}^2\>_{\beta}}\,.
\end{equation}
The integral in (\ref{eq:integral_hermite1}) can be evaluated in
closed form of \cite{GRA07}
\begin{equation}\label{eq:integral_hermite2}
    \textstyle (\text{I})\, =\, \sqrt{\pi}\, \left(1 - s^2\right)^{n/2}\,
    H_n\left(\frac{s\,y}{\sqrt{1 - s^2}}\right)\,.
\end{equation}
For $\phi_n(q)$ to be an eigenstate of the operator $\hat{\rho}_s$,
from equations
(\ref{eq:eigenvalue_problem1})-(\ref{eq:eigenvalue_problem2}) with
(\ref{eq:integral_hermite2}) we need to require the argument of the
Hermite polynomial, $\frac{s\,y}{\sqrt{1 - s^2}} = (v^2 -
\frac{1}{4}) (\tilde{v} \sqrt{\tilde{v}^2 -
2\,\tilde{c}^2\,\<\hat{q}^2\>_{\beta}}\,)^{-1}\, \tilde{c}\,q$ to
equal $\tilde{c}\,q$, which immediately gives
\begin{equation}\label{eq:effective_c}
    \textstyle\tilde{c}\, =\, \left(\frac{\<\hat{p}^2\>_{\beta}}{\hbar^2\,\<\hat{q}^2\>_{\beta}}\right)^{1/4}
\end{equation}
and subsequently the probability for the $n$th eigenstate
$\phi_n(q)$ as
\begin{equation}\label{eq:eigenvalue_problem5}
    \textstyle p_n\, =\, \frac{1}{v + \frac{1}{2}}\,\left(\frac{v - \frac{1}{2}}{v + \frac{1}{2}}\right)^n\,.
\end{equation}
As a result, we obtained the eigenvalues $p_n$ and the eigenstates
$\phi_n(q)$ in closed form.

From comparison between (\ref{eq:parameter_c}) and
(\ref{eq:effective_c}), we introduce an effective mass $M_{\text{\sc
eff}}$ and an effective frequency $\omega_{\text{\sc eff}}$, which
satisfy the relationship
\begin{equation}\label{eq:mass_frequency_rel1}
    \textstyle M_{\text{\sc eff}}\,\omega_{\text{\sc eff}}\, =\,
    \sqrt{\<\hat{p}^2\>_{\beta}/\<\hat{q}^2\>_{\beta}}\,.
\end{equation}
For an uncoupled oscillator, this obviously reduces to
$\<\hat{p}^2\>_{\beta}/\<\hat{q}^2\>_{\beta} = (M \omega_0)^2$. For
a later purpose, it is useful to note that either $M_{\text{\sc
eff}}$ or $\omega_{\text{\sc eff}}$ is not yet determined. The
probability in (\ref{eq:eigenvalue_problem5}) can then be rewritten
as $p_n = (1 - \xi_{\beta}) (\xi_{\beta})^n$ in terms of
$\xi_{\beta} = (v - \frac{1}{2})/(v + \frac{1}{2}) =
e^{-\beta_{\text{\sc eff}}\,\hbar \omega_{\text{\sc eff}}}$ with an
effective temperature
\begin{equation}\label{eq:effective_temperature1}
    \textstyle\beta_{\text{\sc eff}}\, =\, -\frac{\ln \xi_{\beta}}{\hbar\,\omega_{\text{\sc eff}}}\,,
\end{equation}
and subsequently as
\begin{equation}\label{eq:effective_probability1}
    \textstyle p_n\, =\, \frac{1}{Z_{\text{\sc eff}}}\,e^{-\beta_{\text{\sc eff}}\,\hbar \omega_{\text{\sc eff}}\,(n + \frac{1}{2})}
\end{equation}
in terms of an effective partition function $Z_{\text{\sc eff}} =
\sum_n e^{-\beta_{\text{\sc eff}}\,\hbar \omega_{\text{\sc eff}} (n
+ \frac{1}{2})} = \{\text{csch} (\beta_{\text{\sc eff}}\,\hbar
\omega_{\text{\sc eff}}/2)\}/2$. Therefore the density operator
$\hat{\rho}_s$ of the coupled oscillator is equivalent to that
($\hat{\rho}_{\text{\sc eff}}$) of an {\em uncoupled} linear
oscillator
\begin{equation}\label{eq:effective_oscillator1}
    \textstyle\hat{H}_{\text{\sc eff}}\; =\; \frac{\hat{p}^2}{2 M_{\text{\sc eff}}}\, +\,
    \frac{k_{\text{\sc eff}}}{2}\,\hat{q}^2
\end{equation}
at temperature $T_{\text{\sc eff}} = 1/(k_B \beta_{\text{\sc eff}})$
such that $\hat{\rho}_s = e^{-\beta_{\text{\sc eff}}
\hat{H}_{\text{\sc eff}}}/Z_{\text{\sc eff}}$. Here the spring
constant $k_{\text{\sc eff}} = M_{\text{\sc eff}}\,\omega_{\text{\sc
eff}}^2$. Needless to say, the temperature $T_{\text{\sc eff}} \to
T$ if the coupling constants $c_j \to 0$ in
(\ref{eq:total_hamiltonian2}). With the aid of
(\ref{eq:mass_frequency_rel1}) and (\ref{eq:effective_temperature1})
we can easily confirm that
\begin{equation}\label{eq:effective_q_p_sqr}
    \textstyle\<\hat{q}^2\>_{\beta}\, =\, \frac{\hbar}{2 M_{\text{\sc eff}}\,\omega_{\text{\sc eff}}}\,\coth\left(\frac{\beta_{\text{\sc eff}}\,\hbar
    \omega_{\text{\sc eff}}}{2}\right)\; ;\;
    \<\hat{p}^2\>_{\beta}\, =\, \frac{M_{\text{\sc eff}}\,\hbar \omega_{\text{\sc eff}}}{2}\,\coth\left(\frac{\beta_{\text{\sc eff}}\,\hbar
    \omega_{\text{\sc eff}}}{2}\right)\,,
\end{equation}
which can subsequently be used for the internal energy
\begin{equation}\label{eq:effective_internal_energy1}
    \textstyle U_{\text{\sc eff}}\, =\, \<\hat{H}_{\text{\sc eff}}\>_{\beta}\, =\, \frac{\hbar
    \omega_{\text{\sc eff}}}{2}\,\coth \frac{\beta_{\text{\sc eff}}\,\hbar \omega_{\text{\sc eff}}}{2}\, =\, \omega_{\text{\sc eff}} \sqrt{\<\hat{p}^2\>_{\beta}\,\<\hat{q}^2\>_{\beta}}
\end{equation}
(note again, though, that since $\omega_{\text{\sc eff}}$ is not yet
determined, $U_{\text{\sc eff}} \ne U_s$). Further, choosing the
effective frequency
\begin{equation}\label{eq:special_effective_frequency1}
    \textstyle\omega_{\text{\sc eff}}^{\star}\, =\, \frac{1}{2 M}\,\sqrt{\frac{\<\hat{p}^2\>_{\beta}}{\<\hat{q}^2\>_{\beta}}}\, +\,
    \frac{M \omega_0^2}{2}\,\sqrt{\frac{\<\hat{q}^2\>_{\beta}}{\<\hat{p}^2\>_{\beta}}}
\end{equation}
from (\ref{eq:mass_frequency_rel1}), we then have $U_{\text{\sc
eff}}^{\star} = U_s$. Accordingly, $M_{\text{\sc eff}}^{\star} =
\<\hat{p}^2\>_{\beta}/U_s$ and
\begin{equation}\label{eq:special_effective_spring_constant1}
    \textstyle k_{\text{\sc eff}}^{\star}\, =\, \frac{1}{2} \left(k_0\, +\,
    \frac{\<\hat{p}^2\>_{\beta}}{M\,\<\hat{q}^2\>_{\beta}}\right)\,.
\end{equation}
Here $k_0$ is the spring constant of the uncoupled oscillator
$\hat{H}_s$. All effective parameters are now uniquely determined in
terms of the starred quantities, namely, $M_{\text{\sc
eff}}^{\star}$, $k_{\text{\sc eff}}^{\star}$, the effective
temperature $T_{\text{\sc eff}}^{\star} = -\hbar \omega_{\text{\sc
eff}}^{\star}/(k_B\,\ln \xi_{\beta})$, and the internal energy
$U_{\text{\sc eff}}^{\star} = \<\hat{H}_{\text{\sc
eff}}^{\star}\>_{\beta}\, (= U_s)$ where $\hat{H}_{\text{\sc
eff}}^{\star} = \hat{p}^2/(2 M_{\text{\sc eff}}^{\star}) +
(k_{\text{\sc eff}}^{\star}/2)\,\hat{q}^2$ ({\em cf}. note, on the
other hand, that clearly $\hat{\rho}_s = \hat{\rho}_{\text{\sc eff}}
= \hat{\rho}_{\text{\sc eff}}^{\star}$). With the aid of
(\ref{eq:x_drude}) and (\ref{eq:p_drude}), figure \ref{fig:fig1}
demonstrates that $k_{\text{\sc eff}}^{\star} \geq k_0$, which leads
to $M_{\text{\sc eff}}^{\star} \geq M$ from $M_{\text{\sc
eff}}^{\star}\,k_{\text{\sc eff}}^{\star} =
\<\hat{p}^2\>_{\beta}/\<\hat{q}^2\>_{\beta}$. We will use the
effective oscillator with $(M_{\text{\sc eff}}^{\star}, k_{\text{\sc
eff}}^{\star}, T_{\text{\sc eff}}^{\star})$ in
Section~\ref{sec:clausius} for a generalization of the Clausius
inequality.

Let us consider the thermal entropy of the effective uncoupled
oscillator $\hat{H}_{\text{\sc eff}}$ (of course,
$\hat{H}_{\text{\sc eff}}^{\star}$, too, as its special case), which
is
\begin{equation}\label{eq:effective_oscillator_entropy1}
    \textstyle S_{\text{\sc eff}}\; \left(= S_{\text{\sc eff}}^{\star}\right)\, =\, k_B \ln Z_{\text{\sc eff}}\, +\, k_B \beta_{\text{\sc eff}}\,U_{\text{\sc eff}}\, =\, -k_B \left\{\ln (1 - \xi_{\beta})\,
    +\, \frac{\xi_{\beta}}{1 - \xi_{\beta}}\,\ln \xi_{\beta}\right\}\,.
\end{equation}
We can immediately verify that this is identical to the von-Neumann
entropy of the coupled oscillator,
\begin{equation}\label{eq:von-Neumann_entropy1}
    S_N\, =\, \textstyle k_B \left(v + \frac{1}{2}\right)\,
    \ln\left(v + \frac{1}{2}\right) - k_B \left(v - \frac{1}{2}\right)\, \ln\left(v - \frac{1}{2}\right)
\end{equation}
\cite{ALL00} (note again that $\hat{\rho}_s = \hat{\rho}_{\text{\sc
eff}} = \hat{\rho}_{\text{\sc eff}}^{\star}$). From figure
\ref{fig:fig2}, it is shown that $S_N$ increases with the magnitude
of the damping parameter and also with the temperature of the total
system.

Now we briefly comment on the introduction of effective parameters:
First, it has been shown in \cite{SRE93} that the coupled oscillator
with $(M, \omega_0)$ at zero temperature of the total system $(T =
0)$ can be interpreted as an uncoupled one with $(M,
\bar{\omega}_{\text{\sc eff}})$ in a thermal state with a finite
effective temperature $\bar{T}_{\text{\sc eff}}$ (with $\beta =
\infty$), where $\bar{\omega}_{\text{\sc eff}} =
\sqrt{\<\hat{p}^2\>_{\infty}/\<\hat{q}^2\>_{\infty}}/M$. Using the
very same technique, we have generalized this result into
$(M_{\text{\sc eff}}, \omega_{\text{\sc eff}}, T_{\text{\sc eff}})$
in (\ref{eq:mass_frequency_rel1}) and
(\ref{eq:effective_temperature1}) for an arbitrary temperature of
the total system. It is interesting to note here that
$\bar{U}_{\text{\sc eff}} \ne U_s(\beta = \infty)$ though, where
$\bar{U}_{\text{\sc eff}} = \<\hat{p}^2\>_{\infty}/(2M) + \{M
(\bar{\omega}_{\text{\sc eff}})^2/2\} \<\hat{q}^2\>_{\infty}$.
Secondly, it is also known \cite{WEI99,GRA84} that the coupled
oscillator can exactly be seen as an uncoupled oscillator with an
effective frequency
\begin{equation}\label{eq:effective_frequency1}
    \textstyle\tilde{\omega}_{\text{\sc eff}}\, =\, \frac{2}{\hbar \beta}\;
    \mbox{arccoth}\left(\frac{2}{\hbar}\sqrt{\<\hat{q}^2\>_{\beta}\,
    \<\hat{p}^2\>_{\beta}}\right)
\end{equation}
and an effective mass $\tilde{M}_{\text{\sc eff}} =
\sqrt{\<\hat{p}^2\>_{\beta}/\<\hat{q}^2\>_{\beta}}/\tilde{\omega}_{\text{\sc
eff}}$ in the canonical thermal state $\hat{\rho}_{s} = e^{-\beta
\hat{H}_s}/Z_{\beta}$ ({\em i.e.}, $\tilde{T}_{\text{\sc eff}} =
T$), which can be well understood simply as a special case of
$(M_{\text{\sc eff}}, \omega_{\text{\sc eff}}, T_{\text{\sc eff}})$
in (\ref{eq:mass_frequency_rel1}) and
(\ref{eq:effective_temperature1}) (note again that these are not the
starred quantities). However, $\tilde{U}_{\text{\sc eff}} \ne U_s$,
either, whereas $U_s = U_{\text{\sc eff}}^{\star}$ introduced below
equation (\ref{eq:special_effective_frequency1}).

It is also interesting to compare the internal energy $U_{\text{\sc
eff}}^{\star}$ of the coupled oscillator with an alternative
definition \cite{FOR06,KIM06,KIM07,HAE05,FOR05,HAE06}
\begin{equation}\label{eq:alternative_energy1}
    \textstyle{\mathcal U}\, =\, -\frac{\partial}{\partial \beta}\,\ln {\mathcal
    Z}_{\beta}\,,
\end{equation}
where the partition function ${\mathcal Z}_{\beta} =
\mbox{Tr}\,e^{-\beta \hat{H}}/\mbox{Tr}_b\,e^{-\beta \hat{H}_b}$.
Here, $\mbox{Tr}_b$ denotes the partial trace for the bath alone (in
the absence of a coupling between system and bath, the function
${\mathcal Z}_{\beta}$ would exactly correspond to the partition
function of the system only). Then it immediately follows that
${\mathcal U} = U_s + \<\hat{H}_{b}\>_{\beta} +
\<\hat{H}_{sb}\>_{\beta} - \<\hat{H}_{b}\>_{\beta'} \ne U_s$ where
$\<\hat{H}_{b}\>_{\beta'} = \mbox{Tr}_b \left(\hat{H}_b\; e^{-\beta
\hat{H}_b}/\mbox{Tr}_b\,e^{-\beta \hat{H}_b}\right)$. Therefore the
energy ${\mathcal U}$ is not valid for the reduced system alone. The
entropy ${\mathcal S} = k_B\,(\ln {\mathcal Z}_{\beta} -
\beta\,\frac{\partial}{\partial \beta} \ln {\mathcal Z}_{\beta})$
can also be introduced here \cite{FOR05,HAE06}, which is, however,
different from the von-Neumann entropy $S_N\; (= S_{\text{\sc
eff}}^{\star})$ for the reduced density matrix $\hat{\rho}_s$ of the
coupled oscillator. Actually, the entropy ${\mathcal S}$ cannot be
derived from the Jaynes maximum entropy principle \cite{JAY57,EJA57}
applied for the reduced system whereas the entropy $S_{\text{\sc
eff}}^{\star}$ can be so with the effective temperature
$T_{\text{\sc eff}}^{\star}$. As a result, all thermodynamic
quantities resulting from the partition function ${\mathcal
Z}_{\beta}$ are not appropriate for the well-defined {\em local}
thermodynamics of the reduced system.

\section{Clausius inequalities}\label{sec:clausius}
We will discuss the second law of thermodynamics in terms of the
Clausius inequality. To do so, we need the relationship obtained
from (\ref{eq:hamiltonian_internal_energy1}),
\begin{equation}\label{eq:internal_energy_heat_work1}
    d U_s\, =\, \sum_n\textstyle \left(E_n\,dp_n\,+\,p_n\,dE_n\right)
\end{equation}
where $\sum E_n dp_n = \mbox{Tr}_s (\hat{H}_s\,d\hat{\rho}_s) =
\delta{\mathcal Q}_s$ corresponds to an amount of heat added to the
coupled oscillator, and $\sum p_n dE_n = \mbox{Tr}_s
(\hat{\rho}_s\,d\hat{H}_s) = \delta{\mathcal W}_s$ an amount of work
on the oscillator \cite{ALL00}. For a later purpose, we first
consider the well-defined Clausius inequality for a weakly coupled
oscillator
\begin{equation}\label{eq:second_law2_0}
    \textstyle\delta{\mathcal Q}_s \leq T\,dS\,.
\end{equation}
For a typical reversible process we have either a variation of the
mass of the oscillator or a variation of its spring constant in such
a way that
\begin{eqnarray}
    \textstyle\frac{\partial}{\partial M} {\mathcal Q}_s &=&\textstyle \frac{1}{2M}
    \frac{\partial}{\partial M} \<\hat{p}^2\>_{\beta}\, +\,
    \frac{k_0}{2} \frac{\partial}{\partial M}
    \<\hat{q}^2\>_{\beta}\label{eq:uncoupled_second_law1}\\
    &=&\textstyle \frac{\beta\,(\hbar \omega_0)^2}{8M}\,\left(\text{csch} \frac{\beta
    \hbar \omega_0}{2}\right)^2\, =\, T\,\frac{\partial}{\partial M}
    S\\
    \textstyle\frac{\partial}{\partial k_0} {\mathcal Q}_s &=&\textstyle \frac{1}{2M}
    \frac{\partial}{\partial k_0} \<\hat{p}^2\>_{\beta}\, +\,
    \frac{k_0}{2} \frac{\partial}{\partial k_0}
    \<\hat{q}^2\>_{\beta}\label{eq:uncoupled_second_law2}\\
    &=&\textstyle -\frac{\beta\,\hbar^2}{8M}\,\left(\text{csch} \frac{\beta
    \hbar \omega_0}{2}\right)^2\, =\, T\,\frac{\partial}{\partial k_0} S\,,
\end{eqnarray}
respectively ({\em cf}. it is also noted that $\partial {\mathcal
W}_s/\partial M = -\<\hat{p}^2\>_{\beta}/(2 M^2)$ and $\partial
{\mathcal W}_s/\partial k_0 = \<\hat{q}^2\>_{\beta}/2$). We thus
confirm the equality sign in equation (\ref{eq:second_law2_0}).

Next we consider the Clausius inequality in the process of coupling
between oscillator and bath. The coupling process can be represented
in terms of a variation of the damping parameter such that for a
reversible process,
\begin{eqnarray}
    \textstyle\frac{\partial}{\partial \gamma}{\mathcal Q}_s &=&\textstyle \frac{\partial}{\partial
    \gamma} U_s\, =\, \frac{1}{2M} \frac{\partial}{\partial \gamma}
    \<\hat{p}^2\>_{\beta}\, +\, \frac{k_0}{2} \frac{\partial}{\partial \gamma}
    \<\hat{q}^2\>_{\beta}\label{eq:coupling_variation_second_law1}\\
    \textstyle\frac{\partial}{\partial \gamma} S &=&\textstyle k_B
    \left(\frac{\partial v}{\partial \gamma}\right)\, \ln \frac{v + \frac{1}{2}}{v -
    \frac{1}{2}}\label{eq:coupling_variation_entropy1}
\end{eqnarray}
where $S = S_N$ and
\begin{equation}\label{eq:coupling_variation_entropy2}
    \textstyle\frac{\partial v}{\partial \gamma}\, =\, \frac{1}{2 \hbar}
    \frac{\<\hat{q}^2\>_{\beta}\,\frac{\partial}{\partial \gamma}\<\hat{p}^2\>_{\beta}\,
    +\, \<\hat{p}^2\>_{\beta}\,\frac{\partial}{\partial \gamma}\<\hat{q}^2\>_{\beta}}{\sqrt{\<\hat{q}^2\>_{\beta}\,\<\hat{p}^2\>_{\beta}}}\,.
\end{equation}
Equation (\ref{eq:coupling_variation_second_law1}) can be evaluated
in closed form with the aid of equations
(\ref{eq:x_drude})-(\ref{eq:parameter_change0}) such that
\begin{eqnarray}
    \textstyle\frac{\partial}{\partial \gamma} \<\hat{q}^2\>_{\beta}^{(d)} &=&
    {\textstyle \frac{1}{M}} \sum_{l=1}^3 {}_{\gamma}\hspace*{-.05cm}K_l^{(d)}\label{eq:derivative_x_drude_gamma}\\
    \textstyle\frac{\partial}{\partial \gamma} \<\hat{p}^2\>_{\beta}^{(d)}
    &=& -M \sum_{l=1}^3\textstyle\left({}_{\gamma}\hspace*{-.05cm}K_l^{(d)} \underline{\omega_l}^2\,+\,2
    \underline{\omega_l}\,\lambda_d^{(l)}
    \left\{\frac{1}{\beta \underline{\omega_l}}\,+\,\frac{\hbar}{\pi}\; \psi\left(\frac{\beta \hbar
    \underline{\omega_l}}{2\pi}\right)\right\} \frac{\partial \underline{\omega_l}}{\partial \gamma}\right)\label{eq:derivative_x_dot_drude_gamma}
\end{eqnarray}
where $\partial \underline{\omega_1}/\partial \gamma = 0$ and
\begin{equation}
    \textstyle\frac{\partial \underline{\omega_2}}{\partial
    \gamma}\, =\, \frac{1}{2} \left(1 - \frac{1}{\sqrt{1 - \left(\frac{2 {\mathbf
    w}_0}{\gamma}\right)^2}}\right)\; ;\; \frac{\partial \underline{\omega_3}}{\partial \gamma}\,
    =\, \frac{1}{2} \left(1 + \frac{1}{\sqrt{1 - \left(\frac{2 {\mathbf
    w}_0}{\gamma}\right)^2}}\right)\,,\label{eq:drude_model_form_gamma1}
\end{equation}
and
\begin{equation}
    \textstyle {}_{\gamma}\hspace*{-.05cm}K_l^{(d)}\, =\, \left\{\frac{1}{\beta \underline{\omega_l}}\,+\,\frac{\hbar}{\pi}\; \psi\left(\frac{\beta \hbar
    \underline{\omega_l}}{2
    \pi}\right)\right\} \frac{\partial \lambda_d^{(l)}}{\partial
    \gamma}\,+\,\lambda_d^{(l)}\,\left\{-\frac{1}{\beta \underline{\omega_l}^2}\,+\,\frac{\hbar^2 \beta}{2\pi^2}\; \psi^{(1)}(\frac{\beta \hbar
    \underline{\omega_l}}{2\pi})\right\} \frac{\partial \underline{\omega_l}}{\partial \gamma}\label{eq:K_l_gamma}
\end{equation}
in terms of the digamma function $\psi(y)$ and the trigamma function
$\psi^{(1)}(y) = d^2 \ln\Gamma(y)/dy^2$ \cite{ABS74}. Here we
obtained equations (\ref{eq:derivative_x_drude_gamma}) and
(\ref{eq:derivative_x_dot_drude_gamma}) for the overdamped case
$(\gamma/2 > {\mathbf w}_0)$, which is, still, found to hold for the
underdamped case $(\gamma/2 \leq {\mathbf w}_0)$ as well, being
expressed in terms of the functions with complex-valued arguments.
Then we get a violation of the Clausius inequality,
$\partial{\mathcal Q}_s/\partial \gamma
> T\,\partial S_N/\partial \gamma$ as seen from figure \ref{fig:fig3}.

We, however, argue that this violation results from an inappropriate
choice of temperature $T$ being defined for the {\em total} system.
We now propose a well-defined form of the Clausius inequality
pertaining to the coupling process in such a way that
\begin{equation}\label{eq:effective_second_law00}
    \textstyle\delta{\mathcal Q}_{\text{\sc eff}}^{\star}\,\leq\,T_{\text{\sc eff}}^{\star}\,dS_N
\end{equation}
where $\delta{\mathcal Q}_{\text{\sc eff}}^{\star}$ is the heat
exchanged between the effective (weakly coupled) oscillator with
$(M_{\text{\sc eff}}^{\star}, k_{\text{\sc eff}}^{\star})$ in
(\ref{eq:special_effective_spring_constant1}) and a bath at the
equilibrium temperature $T_{\text{\sc eff}}^{\star} = -\hbar
\omega_{\text{\sc eff}}^{\star}/(k_B \ln \xi_{\beta})$ with
(\ref{eq:special_effective_frequency1}). For an reversible process
we then have
\begin{equation}\label{eq:effective_second_law4}
    \textstyle\frac{\partial}{\partial \gamma}{\mathcal Q}_{\text{\sc eff}}^{\star}\, =\, \frac{1}{2 M_{\text{\sc eff}}^{\star}}\,\frac{\partial}{\partial \gamma}
    \<\hat{p}^2\>_{\beta}\,+\,\frac{k_{\text{\sc eff}}^{\star}}{2} \frac{\partial}{\partial \gamma}
    \<\hat{q}^2\>_{\beta}\,,
\end{equation}
which can be shown to be identical to $T_{\text{\sc
eff}}^{\star}\,\partial S_N/\partial \gamma$ with the aid of
(\ref{eq:coupling_variation_entropy1}) and
(\ref{eq:coupling_variation_entropy2}). Therefore, there is no
violation of the Clausius inequality! From $U_{\text{\sc
eff}}^{\star} = U_s$, we note that $\int_0^{\gamma} dU_{\text{\sc
eff}}^{\star} = \int_0^{\gamma} dU_s = U_s(\gamma) - U_0 = \Delta
U_s$, in which $U_0 = (\hbar \omega_0/2) \coth (\beta \hbar
\omega_0/2)$ for an uncoupled oscillator, and $\oint dU_s = \oint
dU_{\text{\sc eff}}^{\star} = 0$, where $\oint = \int_0^{\gamma} +
\int_{\gamma}^0$ represents a cyclic process of the coupling and
decoupling. From the first law $dU_s = \delta{\mathcal Q}_s +
\delta{\mathcal W}_s = \delta{\mathcal Q}_{\text{\sc eff}}^{\star} +
\delta{\mathcal W}_{\text{\sc eff}}^{\star}$ with $\partial{\mathcal
W}_s/\partial \gamma = 0$ and thus $\partial ({\mathcal
Q}_{\text{\sc eff}}^{\star} - {\mathcal Q}_s)/\partial \gamma =
-\partial {\mathcal W}_{\text{\sc eff}}^{\star}/\partial \gamma$, it
also follows that
\begin{equation}\label{eq:effective_second_law_01}
    \textstyle\frac{\partial}{\partial \gamma} {\mathcal Q}_s\,-\,\frac{\partial}{\partial \gamma} {\mathcal
    W}_{\text{\sc eff}}^{\star}\,=\,T_{\text{\sc eff}}^{\star}\,\frac{\partial}{\partial \gamma} S_N
\end{equation}
where, with the aid of
(\ref{eq:special_effective_spring_constant1}),
\begin{eqnarray}\label{eq:effectve_work1}
    \textstyle\frac{\partial}{\partial \gamma} {\mathcal
    W}_{\text{\sc eff}}^{\star} &=&\textstyle -\frac{1}{2 M_{\text{\sc eff}}^{\star}} \left(\frac{\partial M_{\text{\sc eff}}^{\star}}{\partial
    \gamma}\right) \<\hat{p}^2\>_{\beta}\,+\,\frac{1}{2} \left(\frac{\partial k_{\text{\sc eff}}^{\star}}{\partial
    \gamma}\right) \<\hat{q}^2\>_{\beta}\n\\
    &=&\textstyle\frac{1}{4 M} \left(1 - \frac{M k_0\,\<\hat{q}^2\>_{\beta}}{\<\hat{p}^2\>_{\beta}}\right)
    \frac{\partial}{\partial \gamma} \<\hat{p}^2\>_{\beta}\,+\,\frac{k_0}{4} \left(1 -
    \frac{\<\hat{p}^2\>_{\beta}}{M k_0\,\<\hat{q}^2\>_{\beta}}\right) \frac{\partial}{\partial \gamma}
    \<\hat{q}^2\>_{\beta}\,.
\end{eqnarray}
Figure \ref{fig:fig4} shows that $\partial {\mathcal W}_{\text{\sc
eff}}^{\star}/\partial \gamma \leq 0$, which immediately leads to no
violation of the inequality $\partial {\mathcal Q}_s/\partial \gamma
\leq T_{\text{\sc eff}}^{\star}\,\partial S_N/\partial \gamma$. Here
it should be noted that we have appropriately selected the effective
oscillator $\hat{H}_{\text{\sc eff}}^{\star}$ with $(M_{\text{\sc
eff}}^{\star}, \omega_{\text{\sc eff}}^{\star})$ from
(\ref{eq:mass_frequency_rel1}) to introduce an effective temperature
$T_{\text{\sc eff}}^{\star}$ without any ambiguity, which is now a
critical element for the well-defined Clausius inequality in
(\ref{eq:effective_second_law00}).

Therefore, we are now in a position to understand, by means of the
Clausius inequality (\ref{eq:effective_second_law00}), the validity
of the second law in a cyclic process of the coupling and decoupling
between oscillator and bath at an equilibrium temperature $T$. The
validity has actually been shown for zero temperature $(T = 0)$ in
\cite{FOR06,KIM06} and later for an arbitrary temperature in
\cite{KIM07} by verifying the second law in its Kelvin-Planck form
\cite{ANN02}; it states that the minimum work $\Delta{\mathcal F}$
needed to couple the oscillator to a bath (in a reversible process),
being equivalent to the Helmholtz free energy of the {\em coupled}
total system minus the free energy of the {\em uncoupled} total
system, cannot be less than the maximum useful work obtainable from
the oscillator when it decouples from the bath such that
\begin{equation}\label{eq:kim_oconnell1}
    \textstyle\Delta U_s\,\lneq\,\Delta{\mathcal F}
\end{equation}
(note the strict inequality and see below). Here we have on the left
hand side the internal energy $\Delta U_s$ as the maximum useful
work obtainable from the oscillator on completion of the decoupling
process.

For the coupling-decoupling process (with a varying damping
parameter $\gamma': 0 \to \gamma \to 0$), inequality
(\ref{eq:effective_second_law00}) can be transformed to $\oint
\delta{\mathcal Q}_{\text{\sc eff}}^{\star}/T_{\text{\sc
eff}}^{\star} \leq 0$, which means, according to the Kelvin-Planck
form, that the net work obtainable from the effective uncoupled
oscillator (with an accordingly varying parameter $\gamma'$) on
completion of this cyclic process cannot be greater than zero. For a
reversible process, this inequality then reduces to
\begin{equation}\label{eq:effective_second_law31}
    \textstyle\int_0^{\gamma} \frac{d\gamma'}{T_{\text{\sc eff}}^{\star}}\,\frac{\partial}{\partial
    \gamma'} {\mathcal Q}_{\text{\sc eff}}^{\star}\,+\,\int_{\gamma}^0 \frac{d\gamma'}{T_{\text{\sc eff}}^{\star}}\,\frac{\partial}{\partial
    \gamma'} {\mathcal Q}_{\text{\sc eff}}^{\star}\,=\,0\,,
\end{equation}
which means that the minimum work $(\Delta U_s)$ done onto the
oscillator for $U_0 \to U_{\text{\sc eff}}^{\star}$ on completion of
the coupling exactly equals the maximum useful work releasable from
the oscillator on completion of the decoupling. For comparison, on
the other hand, the free energy $\Delta{\mathcal F}$ is, by
definition, the minimum work done on {\em both} oscillator and bath
so that we get the inequality in (\ref{eq:kim_oconnell1}). Anyhow,
the second law holds in the coupling-decoupling process.

Now we consider the Clausius inequality (with a fixed coupling
strength) after completion of the coupling, which has been discussed
so far, e.g., in \cite{ALL00,NIE02,LUT08}. Note that we are now with
the effective oscillator with $(M_{\text{\sc eff}}^{\star},
k_{\text{\sc eff}}^{\star})$ at temperature $T_{\text{\sc
eff}}^{\star}$. We can then show that for a reversible process,
\begin{eqnarray}
    \hspace*{-.3cm}\textstyle\frac{\partial}{\partial M} {\mathcal Q}_{\text{\sc eff}}^{\star} &=& \textstyle\frac{1}{2 M_{\text{\sc eff}}^{\star}}
    \frac{\partial}{\partial M} \<\hat{p}^2\>_{\beta}\,+\,\frac{k_{\text{\sc eff}}^{\star}}{2} \frac{\partial}{\partial M} \<\hat{q}^2\>_{\beta}\,
    =\, \frac{\hbar \omega_{\text{\sc eff}}^{\star}}{4} \left(\frac{\partial}{\partial M} \ln \xi_{\beta}\right)
    \left(\text{csch} \frac{\ln \xi_{\beta}}{2}\right)^2\, =\, T_{\text{\sc eff}}^{\star}\,\frac{\partial}{\partial
    M} S_N\label{eq:effective_second_law0}\\
    \hspace*{-.3cm}\textstyle\frac{\partial}{\partial k_0} {\mathcal Q}_{\text{\sc eff}}^{\star} &=& \textstyle\frac{1}{2 M_{\text{\sc eff}}^{\star}}
    \frac{\partial}{\partial k_0} \<\hat{p}^2\>_{\beta}\,+\,\frac{k_{\text{\sc eff}}^{\star}}{2} \frac{\partial}{\partial k_0}
    \<\hat{q}^2\>_{\beta}\, =\, \frac{\hbar \omega_{\text{\sc eff}}^{\star}}{4} \left(\frac{\partial}{\partial k_0} \ln \xi_{\beta}\right)
    \left(\text{csch} \frac{\ln \xi_{\beta}}{2}\right)^2\, =\, T_{\text{\sc eff}}^{\star}\,\frac{\partial}{\partial k_0} S_N\,,\label{eq:effective_second_law1}
\end{eqnarray}
which follow from (\ref{eq:effective_temperature1}),
(\ref{eq:effective_q_p_sqr}),
(\ref{eq:special_effective_frequency1}) and
(\ref{eq:special_effective_spring_constant1}), respectively.
Therefore, there is no violation of the Clausius inequality at all!
Note here as well that from the first law $dU_s = \delta{\mathcal
Q}_s + \delta{\mathcal W}_s = \delta{\mathcal Q}_{\text{\sc
eff}}^{\star} + \delta{\mathcal W}_{\text{\sc eff}}^{\star}$, the
effective work is also well-defined such as
\begin{equation}\label{eq:effective_work0}
    \textstyle\partial_{M,k_0} {\mathcal W}_{\text{\sc eff}}^{\star}\,
    =\, -\frac{1}{2 M_{\text{\sc eff}}^{\star}} \left(\partial_{M,k_0} M_{\text{\sc
    eff}}^{\star}\right) \<\hat{p}^2\>_{\beta}\,+\,\frac{1}{2} \left(\partial_{M,k_0} k_{\text{\sc
    eff}}^{\star}\right) \<\hat{q}^2\>_{\beta}\,,
\end{equation}
the closed form of which can immediately be obtained with the aid of
$M_{\text{\sc eff}}^{\star} = \<\hat{p}^2\>_{\beta}/U_s$ and
(\ref{eq:special_effective_spring_constant1}).

For comparison, we next seek to have equations
(\ref{eq:uncoupled_second_law1}) and
(\ref{eq:uncoupled_second_law2}) in closed form in the Drude damping
model. After making some lengthy calculations ({\em cf}. Appendix
\ref{sec:appendix3}), we then obtain the expressions
\begin{eqnarray}
    \textstyle\frac{\partial}{\partial M} \<\hat{q}^2\>_{\beta}^{(d)} &=&
    {\textstyle -\frac{1}{M} \<\hat{q}^2\>_{\beta}^{(d)}\,+\,\frac{1}{M}}
    \sum_{l=1}^3 {}_{M}\hspace*{-.05cm}K_l^{(d)}\label{eq:derivative_x_drude}\\
    \textstyle\frac{\partial}{\partial M} \<\hat{p}^2\>_{\beta}^{(d)}
    &=& {\textstyle \frac{1}{M} \<\hat{p}^2\>_{\beta}^{(d)}} - M \sum_{l=1}^3\textstyle\left({}_{M}\hspace*{-.05cm}K_l^{(d)} \underline{\omega_l}^2\,+\,2
    \underline{\omega_l}\,\lambda_d^{(l)}
    \left\{\frac{1}{\beta \underline{\omega_l}}\,+\,\frac{\hbar}{\pi}\; \psi\left(\frac{\beta \hbar
    \underline{\omega_l}}{2\pi}\right)\right\} \frac{\partial \underline{\omega_l}}{\partial
    M}\right)\label{eq:derivative_x_dot_drude}
\end{eqnarray}
where
\begin{equation}\label{eq:K_l}
    \textstyle {}_{M}\hspace*{-.05cm}K_l^{(d)} := \left\{\frac{1}{\beta \underline{\omega_l}}\,+\,\frac{\hbar}{\pi}\; \psi\left(\frac{\beta \hbar
    \underline{\omega_l}}{2
    \pi}\right)\right\} \frac{\partial \lambda_d^{(l)}}{\partial
    M}\,+\,\lambda_d^{(l)}\,\left\{-\frac{1}{\beta \underline{\omega_l}^2}\,+\,\frac{\hbar^2 \beta}{2\pi^2}\; \psi^{(1)}(\frac{\beta \hbar
    \underline{\omega_l}}{2\pi})\right\} \frac{\partial \underline{\omega_l}}{\partial
    M}\,,
\end{equation}
and
\begin{eqnarray}
    \textstyle\frac{\partial}{\partial k_0} \<\hat{q}^2\>_{\beta}^{(d)} &=&
    {\textstyle \frac{1}{M}} \sum_{l=1}^3 {}_{k_0}\hspace*{-.05cm}K_l^{(d)}\label{eq:derivative_x_drude_k}\\
    \textstyle\frac{\partial}{\partial k_0} \<\hat{p}^2\>_{\beta}^{(d)}
    &=& -M \sum_{l=1}^3\textstyle\left({}_{k_0}\hspace*{-.05cm}K_l^{(d)} \underline{\omega_l}^2\,+\,2
    \underline{\omega_l}\,\lambda_d^{(l)}
    \left\{\frac{1}{\beta \underline{\omega_l}}\,+\,\frac{\hbar}{\pi}\; \psi\left(\frac{\beta \hbar
    \underline{\omega_l}}{2\pi}\right)\right\} \frac{\partial \underline{\omega_l}}{\partial k_0}\right)\label{eq:derivative_x_dot_drude_k}
\end{eqnarray}
where ${}_{k_0}\hspace*{-.05cm}K_l^{(d)}$ is defined as equation
(\ref{eq:K_l}) but with the replacement of $\partial/\partial M$ by
$\partial/\partial k_0$. Equations (\ref{eq:derivative_x_drude}) and
(\ref{eq:derivative_x_dot_drude}) (as well as
(\ref{eq:derivative_x_drude_k}) and
(\ref{eq:derivative_x_dot_drude_k})) hold for both overdamped and
underdamped cases. To discuss the second law, we now consider an
equality similar to (\ref{eq:effective_second_law_01}) in form of
\begin{eqnarray}
    \textstyle\frac{\partial}{\partial M} {\mathcal Q}_s\,-\,\frac{\partial}{\partial M}
    \left({\mathcal W}_{\text{\sc eff}}^{\star} - {\mathcal W}_s\right) &=&\textstyle T_{\text{\sc eff}}^{\star}\,\frac{\partial}{\partial
    M} S_N\label{eq:second_law_well_defined1}\\
    \textstyle\frac{\partial}{\partial k_0} {\mathcal Q}_s\,-\,\frac{\partial}{\partial k_0}
    \left({\mathcal W}_{\text{\sc eff}}^{\star} - {\mathcal W}_s\right) &=&\textstyle T_{\text{\sc eff}}^{\star}\,\frac{\partial}{\partial
    k_0} S_N\label{eq:second_law_well_defined2}
\end{eqnarray}
obtained with the aid of the first law. Here $\partial {\mathcal
W}_s/\partial M = -\<\hat{p}^2\>_{\beta}/(2 M^2)$ and $\partial
{\mathcal W}_s/\partial k_0 = \<\hat{q}^2\>_{\beta}/2$, whereas
$\partial {\mathcal W}_s/\partial \gamma$ for
(\ref{eq:effective_second_law_01}) vanishes. Actually we have here
$\partial_{M,k_0} ({\mathcal W}_{\text{\sc eff}}^{\star} - {\mathcal
W}_s) \geq 0$, which follows from $\partial {\mathcal Q}_s/\partial
M \geq T_{\text{\sc eff}}^{\star}\,\partial S_N/\partial M$ and
$\partial {\mathcal Q}_s/\partial k_0 \geq T_{\text{\sc
eff}}^{\star}\,\partial S_N/\partial k_0$, demonstrated in figures
\ref{fig:fig5} and \ref{fig:fig6}, respectively (where equations
(\ref{eq:derivative_x_drude})-(\ref{eq:derivative_x_dot_drude_k})
were used). This can be interpreted as follows. To define a
well-defined (effective) local temperature of the oscillator, we
need to ``project'' the coupled oscillator onto the effective
oscillator. In doing so, it is required here to do additional work
$({\mathcal W}_{\text{\sc eff}}^{\star} - {\mathcal W}_s)$ onto the
oscillator, whereas in the coupling process we need to release the
work ${\mathcal W}_{\text{\sc eff}}^{\star}$ from the oscillator.
Without considering this work compensation we would consequently get
a violation of the Clausius inequality. It is also interesting to
rewrite (\ref{eq:second_law_well_defined1}) as
\begin{equation}\label{eq:second_law2_21}
    \textstyle\frac{\partial}{\partial M} {\mathcal Q}_s\,=\,T\,\frac{\partial}{\partial M}
    S_N\,+\,Y
\end{equation}
in terms of the temperature $T$ of the {\em total} system, where $Y
= (T_{\text{\sc eff}}^{\star} - T)\,\partial S_N/\partial
M\,+\,\partial ({\mathcal W}_{\text{\sc eff}}^{\star} - {\mathcal
W}_s)/\partial M$. Figure \ref{fig:fig7} shows that $\partial
{\mathcal Q}_s/\partial M > T\,\partial S_N/\partial M$, which has
been used in \cite{LUT08} for the justification of a violation of
the Clausius inequality (note also the strict inequality even for a
reversible process). However, we understand now that this simply
represents the neglect of the additional term $Y > 0$ rather than a
violation proper. As a result, we have a generalized form of the
Clausius inequality, $\oint \delta{\mathcal Q}_{\text{\sc
eff}}^{\star}/T_{\text{\sc eff}}^{\star} \leq 0$ where $\oint$
represents a cyclic process with respect to {\em any} variation of
$(M, k_0, \gamma)$. For the relevant comment on effective
thermodynamic relations, see Appendix \ref{sec:comment2}.

\section{Conclusion}\label{sec:conclusion}
In conclusion, we have found a well-defined effective Clausius
inequality appropriate for the quantum Brownian oscillator with {\em
any} coupling strength. It satisfies the equality condition for a
reversible process. We have clearly shown that there is no violation
of the inequality so that the second law of thermodynamics is robust
even beyond the weak coupling limit. In doing so, we have used the
effective internal energy $U_{\text{\sc eff}}^{\star} =
\<\hat{H}_{\text{\sc eff}}^{\star}\>_{\beta}$, being identical to
the internal energy $U_s = \<\hat{H}_s\>_{\beta}$, whereas the
approach of apparently many other works has been based on a
different energy ${\mathcal U}$ defined in
(\ref{eq:alternative_energy1}) and discussed thereafter.

We believe that this inequality will provide a useful starting point
for a consistent generalization of thermodynamics and information
theory into the quantum and nano-system regime, respectively. As an
example, a generalization of the Landauer principle
\cite{LAN61,BEN03} is in consideration, which can be understood as a
simple logical consequence of the Clausius inequality; our findings
then suggest an existence of an effective Landauer principle yet to
be introduced rigorously \cite{KIM09} which is correct even in the
strong coupling limit, whereas based on the violation of the
Clausius inequality considered in \cite{ALL00,NIE02} as stated in
Section \ref{sec:introduction}, it was concluded in \cite{BUE08}
that the original form of Landauer principle may not be applicable
in the strong coupling limit.

We thank A.E. Allahverdyan for his critical reading of the
manuscript. One of us (I.K.) is grateful to Prof. Jaewan Kim at KIAS
for bringing Ref. \cite{SRE93} into his attention. We also
appreciate all comments and constructive questions of the referees
which made us clarify the paper and improve its quality.

\appendix\section{Derivation of equations (\ref{eq:prob01_zero})-(\ref{eq:prob11}) and (\ref{eq:prob20})-(\ref{eq:prob21})}
\label{sec:appendix1}
Substituting into (\ref{eq:wahrscheinlichkeit1}) the eigenfunction
\begin{equation}\label{eq:eigenfunction1}
    \textstyle \psi_{\nu}(q)\; =\; \sqrt{\frac{c}{2^{\nu} {\nu}! \sqrt{\pi}}}\; e^{-\frac{c^2}{2} q^2}\, H_{\nu}\left(c q\right)
\end{equation}
in terms of the Hermite polynomial $H_{\nu}$ where $\nu = n, m$,
then we immediately obtain
\begin{eqnarray}\label{eq:wahrscheinlichkeit1_1}
    \rho_{nm} &=& \textstyle\frac{1}{\sqrt{2^{n+m}\,n!\,m!}} \frac{1}{\pi\,c} \frac{1}{\sqrt{2\,\<\hat{q}^2\>_{\beta}}} \int_{-\infty}^{\infty}\int_{-\infty}^{\infty} dy\, dy'\, H_n(y)\,
    H_m(y')\, \times\n\\
    && \textstyle\exp\left\{-a_{\beta} (y^2 + y'^2) + \frac{1}{c^2}\left(\frac{\<\hat{p}^2\>_{\beta}}{\hbar^2} - \frac{1}{4 \<\hat{q}^2\>_{\beta}}\right) y\,y'\right\}
\end{eqnarray}
where $y = c q$ and $y' = c q'$, and $a_{\beta} = \frac{1}{2} +
\frac{1}{8 c^2\,\<\hat{q}^2\>_{\beta}} +
\frac{\<\hat{p}^2\>_{\beta}}{2 \hbar^2 c^2}$. The substitution of
the relations \cite{ABS74}
\begin{subequations}
\begin{eqnarray}
    \textstyle H_n(y) &=& \textstyle \left(\frac{\partial}{\partial
    t}\right)^n\,e^{2 y t - t^2}|_{t=0}\\
    \textstyle H_m(y') &=& \textstyle \left(\frac{\partial}{\partial
    s}\right)^m\,e^{2 y' s - s^2}|_{s=0}\label{eq:hermite_polynomial0}
\end{eqnarray}
\end{subequations}
into (\ref{eq:wahrscheinlichkeit1_1}) subsequently allows us to have
\begin{eqnarray}\label{eq:wahrscheinlichkeit2}
    \rho_{nm} &=& \textstyle \frac{1}{\sqrt{2^{n+m}\,n!\,m!}} \frac{1}{\pi\,c} \frac{1}{\sqrt{2\,\<\hat{q}^2\>_{\beta}}} \left(\frac{\partial}{\partial t}\right)^n\left(\frac{\partial}{\partial
    s}\right)^m e^{-(t^2+s^2)}\; \times\n\\
    && \textstyle\left.\int_{-\infty}^{\infty}\int_{-\infty}^{\infty} dy\,dy'\,e^{-(a_{\beta}\,y^2 - 2 t y)}\,e^{-\{a_{\beta}\,y'^2 + 2\,b_{\beta}(y)\cdot y'\}}\right|_{t,s=0}
\end{eqnarray}
where $b_{\beta}(y) = \frac{y}{2 c^2} \left(\frac{1}{4
\<\hat{q}^2\>_{\beta}} -
\frac{\<\hat{p}^2\>_{\beta}}{\hbar^2}\right) - s$. By using the
identity \cite{ABS74}
\begin{equation}\label{eq:integral1}
    \textstyle \int_{-\infty}^{\infty} dz\, e^{-(a z^2 + 2 b z)}\;
    =\; \sqrt{\frac{\pi}{a}}\, e^{\frac{b^2}{a}}
\end{equation}
we can first carry out the integration over $y'$ in
(\ref{eq:wahrscheinlichkeit2}) and then over $y$, which will give
rise to
\begin{eqnarray}\label{eq:wahrscheinlichkeit3}
    \rho_{nm} &=&\textstyle \frac{1}{\sqrt{2^{n+m}\,n!\,m!}} \frac{1}{c} \frac{1}{\sqrt{2\,\<\hat{q}^2\>_{\beta} A_{\beta}}}
    \left(\frac{\partial}{\partial t}\right)^n
    \left(\frac{\partial}{\partial s}\right)^m \left.e^{-\Upsilon_{\beta} (t^2 + s^2) + 2 \Lambda_{\beta}\,t\,s}\right|_{t,s=0}\n\\
    &=&\textstyle \sqrt{\frac{\left(\Upsilon_{\beta}\right)^{m-n}}{2^{n+m}\,n!\,m!}} \frac{\left(\Lambda_{\beta}\right)^n}{c}
    \frac{1}{\sqrt{2\,\<\hat{q}^2\>_{\beta} A_{\beta}}}
    \left.\left(\frac{\partial}{\partial u}\right)^n e^{-\Delta_{\beta}\,u^2}\, H_m(u)\right|_{u=0}
\end{eqnarray}
in terms of $A_{\beta}$, $\Upsilon_{\beta}$, $\Lambda_{\beta}$ and
$\Delta_{\beta}$ in (\ref{eq:prob2}). Here we used equation
(\ref{eq:hermite_polynomial0}), and $u =
\frac{\Lambda_{\beta}\,t}{\sqrt{\Upsilon_{\beta}}}$. From the
Heisenberg uncertainty relation with $\<\hat{q}\>_{\beta} =
\<\hat{p}\>_{\beta} = 0$ ({\em cf}. (\ref{eq:expectation_value1})),
it follows that $1
> \Lambda_{\beta} \geq 0$. For a later purpose it is useful to
confirm that for an uncoupled oscillator, $\Upsilon_{\beta} =
\Delta_{\beta} = 0$ and $\Lambda_{\beta} = e^{-\beta \hbar
\omega_0}$.

To arrive at a closed form for $\rho_{nm}$, we consider the
expression in (\ref{eq:wahrscheinlichkeit3})
\begin{eqnarray}\label{eq:wahrscheinlichkeit5}
    {}_{nm}\hspace*{-.03cm}\Xi_{\beta} &:=& \textstyle\left.\left(\frac{\partial}{\partial u}\right)^n
    e^{-\Delta_{\beta}\,u^2}\, H_m(u)\right|_{u=0}\n\\
    &=& \sum_{r=0}^n \binom{n}{r}\textstyle\left.\left(e^{-\Delta_{\beta}\,u^2}\right)^{(n-r)}\left\{H_m(u)\right\}^{(r)}\right|_{u=0}
\end{eqnarray}
where $\tbinom{n}{r} = \frac{n!}{(n-r)!\,r!}$, and $(\cdots)^{(r)} =
\left(\frac{\partial}{\partial u}\right)^r (\cdots)$. The Hermite
polynomial $H_m(u) = {}_{m}\hspace*{-.03cm}h_m\,u^m +
{}_{m}\hspace*{-.03cm}h_{m-2}\,u^{m-2} + \cdots$ can be expressed as
\cite{ABS74}
\begin{subequations}
\begin{eqnarray}
    H_{2{\mathit l}}(u) &=& \textstyle (-1)^{\mathit l}\,\frac{(2{\mathit l})!}{{\mathit l}!}\; {}_{1}\hspace*{-.07cm}F_{1}\left(-{\mathit l}; \frac{1}{2}; u^2\right)\label{eq:hermite_polynomial1}\\
    H_{2{\mathit l}+1}(u) &=& \textstyle (-1)^{\mathit l}\,\frac{2\cdot(2{\mathit l}+1)!}{{\mathit l}!}\, u\; {}_{1}\hspace*{-.07cm}F_{1}\left(-{\mathit l}; \frac{3}{2}; u^2\right)\label{eq:hermite_polynomial1_1}
\end{eqnarray}
\end{subequations}
in terms of the confluent hypergeometric function
${}_{1}\hspace*{-.07cm}F_1(a; b; z) = \frac{\Gamma(b)}{\Gamma(a)}
\sum_{k=0}^{\infty} \frac{\Gamma(a+k)}{\Gamma(b+k)} \frac{z^k}{k!}$
with $\frac{1}{\Gamma(-k)} = 0$, $k = 0, 1, 2, \cdots$. Then it
immediately follows that
\begin{equation}\label{eq:wahrscheinlichkeit6}
    \textstyle \left.\{H_{2{\mathit l}}(u)\}^{(r)}\right|_{u=0} = \left\{
    \begin{array}{cc}
        (2p)!\,{}_{2{\mathit l}}\hspace*{-.03cm}h_{2p}\;\; &\text{for}\;\; r = 2p\;\; \text{even, where}\;\; p \leq {\mathit l}\\
        0&\text{otherwise}
    \end{array}\right.
\end{equation}
and from (\ref{eq:hermite_polynomial1}),
\begin{equation}\label{eq:hermite_polynomial3}
    \textstyle \left.\{H_{2{\mathit l}}(u)\}^{(2p)}\right|_{u=0}\, =\, \sqrt{\pi}\,(-1)^{\mathit l}
    \frac{(2{\mathit l})!}{{\mathit l}!}\frac{(2p)!}{p!}\frac{(-{\mathit l})_p}{\Gamma\left(p+\frac{1}{2}\right)}
\end{equation}
where the Pochhammer symbol $(z)_k = \frac{\Gamma(z+k)}{\Gamma(z)}$.
Similarly, we can also find that
\begin{equation}\label{eq:wahrscheinlichkeit7}
    \textstyle \left.\{H_{2{\mathit l}+1}(u)\}^{(r)}\right|_{u=0} = \left\{
    \begin{array}{cc}
        (2p+1)!\,{}_{2{\mathit l}+1}\hspace*{-.03cm}h_{2p+1}\;\; &\text{for}\;\; r = 2p+1\;\; \text{odd, where}\;\; p \leq {\mathit l}\\
        0&\text{otherwise}
    \end{array}\right.
\end{equation}
and from (\ref{eq:hermite_polynomial1_1}),
\begin{equation}\label{eq:hermite_polynomial3_1}
    \textstyle\left.\{H_{2{\mathit l}+1}(u)\}^{(2p+1)}\right|_{u=0}\,
    =\, \sqrt{\pi}\,(-1)^{\mathit l} \frac{(2{\mathit l}+1)!}{{\mathit l}!}\frac{(2p+1)!}{p!}\frac{(-{\mathit l})_p}{\Gamma\left(p+\frac{3}{2}\right)}\,.
\end{equation}
Also, in (\ref{eq:wahrscheinlichkeit5}) we have
\begin{equation}\label{eq:wahrscheinlichkeit7_1}
    \textstyle\left.\left(e^{-\Delta_{\beta}\,u^2}\right)^{(n-r)}\right|_{u=0} = \left\{
    \begin{array}{cc}
    \frac{(-\Delta_{\beta})^q (2q)!}{q!}\;\; &\text{for}\;\; n-r = 2q\;\; \text{even}\\
        0&\text{otherwise.}
    \end{array}\right.
\end{equation}
With the aid of
(\ref{eq:wahrscheinlichkeit6})-(\ref{eq:wahrscheinlichkeit7_1}),
equation (\ref{eq:wahrscheinlichkeit5}) reduces to
\begin{eqnarray}
    {}_{n,2{\mathit l}+1}\hspace*{-.03cm}\Xi_{\beta} &=& \textstyle 0\label{eq:wahrscheinlichkeit80}\\
    {}_{n,2{\mathit l}}\hspace*{-.03cm}\Xi_{\beta} &=& \textstyle \frac{\sqrt{\pi}\,(-1)^{k+{\mathit l}}\,(2k)!\,(2{\mathit l})!\,(\Delta_{\beta})^k}{{\mathit l}!\; \Gamma(-{\mathit l})}
    {\displaystyle \sum_{p=0}^k}\,\frac{\Gamma(-{\mathit l}+p)}{\Gamma\left(\frac{1}{2}+p\right)\,\Gamma(k-p+1)}
    \frac{\left(\frac{-1}{\Delta_{\beta}}\right)^p}{p!}\label{eq:wahrscheinlichkeit801}\\
    &=& \textstyle (-1)^{k+{\mathit l}}\,\frac{(2k)!}{k!}\,\frac{(2{\mathit l})!}{{\mathit l}!}\,(\Delta_{\beta})^k\, {}_{2}\hspace*{-.05cm}F_1\left(-k,-{\mathit l}; \frac{1}{2};
    \frac{1}{\Delta_{\beta}}\right)\label{eq:wahrscheinlichkeit8}
\end{eqnarray}
for $n=2k$ even, and similarly
\begin{eqnarray}
    {}_{n,2{\mathit l}}\hspace*{-.03cm}\Xi_{\beta} &=& \textstyle 0\label{eq:wahrscheinlichkeit8_01}\\
    {}_{n,2{\mathit l}+1}\hspace*{-.03cm}\Xi_{\beta} &=& \textstyle 2\,(-1)^{k+{\mathit l}}\,\frac{(2k+1)!}{k!}\,\frac{(2{\mathit l}+1)!}{{\mathit l}!}\,(\Delta_{\beta})^k\,
    {}_{2}\hspace*{-.05cm}F_1\left(-k,-{\mathit l}; \frac{3}{2};
    \frac{1}{\Delta_{\beta}}\right)\label{eq:wahrscheinlichkeit8_1}
\end{eqnarray}
for $n=2k+1$ odd. In (\ref{eq:wahrscheinlichkeit801}) we used the
identity
\begin{equation}\label{eq:relation0}
    \textstyle \frac{1}{\Gamma(k-p+1)}\; =\;
    \frac{(-1)^p\,\Gamma(-k+p)}{\Gamma(k+1)\,\Gamma(-k)}
\end{equation}
to get the hypergeometric function ${}_{2}\hspace*{-.05cm}F_1(a,b;
c; z) = \frac{\Gamma(c)}{\Gamma(a) \Gamma(b)} \sum_{k=0}^{\infty}
\frac{\Gamma(a+k) \Gamma(b+k)}{\Gamma(c+k)} \frac{z^k}{k!}$ in
(\ref{eq:wahrscheinlichkeit8}). From (\ref{eq:wahrscheinlichkeit3}),
(\ref{eq:wahrscheinlichkeit5}) and
(\ref{eq:wahrscheinlichkeit80})-(\ref{eq:wahrscheinlichkeit8_1})
with the relation \cite{ABS74}
\begin{equation}\label{eq:gamma_identity1}
    \textstyle \Gamma(2\nu)\, =\, \frac{1}{\sqrt{\pi}}\, 2^{2\nu-1}\, \Gamma(\nu)\, \Gamma\left(\nu + \frac{1}{2}\right)
\end{equation}
where $\nu = k,{\mathit l}$, we finally get equations
(\ref{eq:prob01_zero})-(\ref{eq:prob11}). We will below simplify the
closed forms in (\ref{eq:prob1}) and (\ref{eq:prob11}),
respectively.

Now we use the relation \cite{ABS74}
\begin{equation}\label{eq:jacobi_identity1}
    \textstyle P_n^{(\mu,\nu)}(z) = \binom{n+\mu}{n}\left(\frac{1+z}{2}\right)^n\,{}_{2}\hspace*{-.05cm}F_1\left(-n,-n-\nu; \mu+1; \frac{z-1}{z+1}\right)
\end{equation}
to express the matrix elements $\left(\hat{\rho}_s\right)_{nm}$ in
terms of the Jacobi polynomial \cite{ABS74}
\begin{equation}\label{eq:jacobi_polynomial1}
    \textstyle P_n^{(\mu,\nu)}(z)\, =\, \frac{1}{2^n} {\displaystyle \sum_{k=0}^n} \binom{n+\mu}{k}\,\binom{n+\nu}{n-k}\,(z-1)^{n-k}\,(z+1)^k
\end{equation}
where $\mu, \nu > -1$. Equation (\ref{eq:jacobi_identity1}) allows
us to have
\begin{subequations}
\begin{eqnarray}
    \textstyle {}_{2}\hspace*{-.05cm}F_1\left(-k,-{\mathit l}; \frac{1}{2}; \frac{1}{\Delta_{\beta}}\right) &=&
    \textstyle\frac{\Gamma(\frac{1}{2})\,\Gamma({\mathit l}+1)}{\Gamma({\mathit l}+\frac{1}{2})} \left(\frac{2}{1+v}\right)^{\mathit l}\,P_{\mathit l}^{(-\frac{1}{2},k-{\mathit l})}(v)\label{eq:jacobi_polynomial20}\\
    \textstyle {}_{2}\hspace*{-.05cm}F_1\left(-k,-{\mathit l}; \frac{3}{2}; \frac{1}{\Delta_{\beta}}\right) &=&
    \textstyle\frac{\Gamma(\frac{3}{2})\,\Gamma({\mathit l}+1)}{\Gamma({\mathit l}+\frac{3}{2})} \left(\frac{2}{1+v}\right)^{\mathit l}\,P_{\mathit l}^{(\frac{1}{2},k-{\mathit l})}(v)\label{eq:jacobi_polynomial21}
\end{eqnarray}
\end{subequations}
for $k \geq {\mathit l}$ where $v = 1 - \frac{2}{1-\Delta_{\beta}}$.
Further, we find that \cite{SZE39}
\begin{subequations}
\begin{eqnarray}
    \textstyle P_{2n}^{(\nu,\nu)}(z) &=& \textstyle
    (-1)^{n}\,\frac{\Gamma(2n+\nu+1)\,\Gamma(n+1)}{\Gamma(n+\nu+1)\,\Gamma(2n+1)}\,P_{n}^{(-\frac{1}{2},\nu)}\left(1 - 2
    z^2\right)\label{eq:jacobi_identities1}\\
    \textstyle P_{2n+1}^{(\nu,\nu)}(z) &=& \textstyle
    (-1)^{n}\,\frac{\Gamma(2n+\nu+2)\,\Gamma(n+1)}{\Gamma(n+\nu+1)\,\Gamma(2n+2)}\,z\,P_{n}^{(\frac{1}{2},\nu)}\left(1 - 2
    z^2\right)\,,\label{eq:jacobi_identities2}
\end{eqnarray}
\end{subequations}
which can be verified, respectively, by using equation
(\ref{eq:jacobi_polynomial1}) and then comparing each coefficient of
$z^k$ on both sides. Note here that $P_{2n}^{(\nu,\nu)}(z) =
P_{2n}^{(\nu,\nu)}(-z)$ and $P_{2n+1}^{(\nu,\nu)}(z) =
-P_{2n+1}^{(\nu,\nu)}(-z)$. With the aid of
(\ref{eq:jacobi_polynomial20})-(\ref{eq:jacobi_identities2}),
equations (\ref{eq:prob1}) and (\ref{eq:prob11}) can then be
transformed into (\ref{eq:prob20}) and (\ref{eq:prob21}),
respectively.

\section{Derivation of equations (\ref{eq:derivative_x_drude})-(\ref{eq:derivative_x_dot_drude_k})}\label{sec:appendix3}
Using the relations in (\ref{eq:parameter_change0}) with $\partial
\omega_0/\partial M = -\omega_0/(2M)$ we can easily obtain
\begin{equation}\label{eq:derivative_drude1}
    \textstyle\frac{\partial \Omega}{\partial M}\, =\, \frac{\Omega}{\Omega +
    \gamma} \frac{{\mathbf w}_0^2\,\gamma}{M\,\{\Omega (\gamma - \Omega) \,-\,{\mathbf
    w}_0^2\}}\; ;\; \frac{\partial \gamma}{\partial M}\, =\, -\frac{\partial \Omega}{\partial
    M}\; ;\; \frac{\partial {\mathbf w}_0}{\partial M}\, =\, \frac{{\mathbf w}_0\,\Omega\,({\mathbf
    w}_0^2\,+\,\Omega^2\,-\,\gamma^2)}{2 M\,(\Omega + \gamma)\,\{\Omega\,(\gamma - \Omega)\,-\,{\mathbf
    w}_0^2\}}\,.
\end{equation}
From $\partial \omega_0/\partial k_0 = -(\partial \omega_0/\partial
M)/\omega_0^2$, it immediately follows as well that
\begin{equation}\label{eq:derivative_drude10}
    \textstyle\frac{\partial \Omega}{\partial k_0}\, =\, -\frac{1}{({\mathbf w}_0)^2} \frac{\Omega + \gamma}{\Omega}\,\frac{\partial \Omega}{\partial M}\; ;\; \frac{\partial \gamma}{\partial k_0}\, =\, -\frac{\partial \Omega}{\partial
    k_0}\; ;\; \frac{\partial {\mathbf w}_0}{\partial k_0}\, =\, -\frac{1}{({\mathbf w}_0)^2} \frac{\Omega + \gamma}{\Omega}\,\frac{\partial {\mathbf w}_0}{\partial M}\,.
\end{equation}
We also have $\partial \underline{\omega_1}/\partial M = \partial
\Omega/\partial M$ and
\begin{eqnarray}\label{eq:drude_model_form1}
    \textstyle\frac{\partial \underline{\omega_2}}{\partial M} &=&
    \textstyle\frac{1}{\sqrt{\left(\frac{\gamma}{2 {\mathbf w}_0}\right)^2 - 1}} \frac{\partial {\mathbf w}_0}{\partial M} -
    \frac{1}{2} \left(1 - \frac{1}{\sqrt{1 - \left(\frac{2 {\mathbf w}_0}{\gamma}\right)^2}}\right) \frac{\partial \Omega}{\partial
    M}\\
    \textstyle\frac{\partial \underline{\omega_3}}{\partial M} &=&
    \textstyle-\frac{1}{\sqrt{\left(\frac{\gamma}{2 {\mathbf w}_0}\right)^2 - 1}} \frac{\partial {\mathbf w}_0}{\partial M} -
    \frac{1}{2} \left(1 + \frac{1}{\sqrt{1 - \left(\frac{2 {\mathbf w}_0}{\gamma}\right)^2}}\right) \frac{\partial \Omega}{\partial
    M}\,.
\end{eqnarray}
And from (\ref{eq:coefficients}),
\begin{equation}\label{eq:drude_model_form2}
    \textstyle\frac{\partial \lambda_d^{(1)}}{\partial M} = \frac{\left(2\,\Omega \gamma - \gamma^2 + \Omega^2 + {\mathbf
    w}_0^2\right)\,\frac{\partial \Omega}{\partial M}\,+\,2\,\gamma {\mathbf w}_0 \frac{\partial {\mathbf w}_0}{\partial M}}{(\Omega -
    z_1)^2\,(\Omega - z_2)^2}\; ;\; \frac{\partial \lambda_d^{(2)}}{\partial M} = \frac{(z_2^2 -
    z_1^2)\,\frac{\partial \Omega}{\partial M}\,+\,(\Omega + z_2) (2\,z_1 - z_2 - \Omega)\,\frac{\partial z_1}{\partial M}\,+\,(\Omega^2 -
    z_1^2)\,\frac{\partial z_2}{\partial M}}{(z_1 - \Omega)^2\,(z_1 -
    z_2)^2}
\end{equation}
and $\partial \lambda_d^{(3)}/\partial M \to \partial
\lambda_d^{(2)}/\partial M$ with ($z_1 \leftrightarrow z_2$). Then
we finally arrive at the expressions in
(\ref{eq:derivative_x_drude}) and (\ref{eq:derivative_x_dot_drude}),
respectively. With the replacement of $\partial/\partial M$ in
(\ref{eq:drude_model_form1})-(\ref{eq:drude_model_form2}) by
$\partial/\partial k_0$, we can also have
(\ref{eq:derivative_x_drude_k}) and
(\ref{eq:derivative_x_dot_drude_k}).

\section{Comment on effective thermodynamic relations}\label{sec:comment2}
In ordinary thermodynamics, the notion of temperature $T$ appears
conceptually as a partial derivative of internal energy $U$ with
respect to entropy $S$ such that $T = \partial U/\partial S$. This
also holds for the effective temperature $T_{\text{\sc
eff}}^{\star}$\,: Combining the first law $d U_s = \delta{\mathcal
Q}_{\text{\sc eff}}^{\star} + \delta{\mathcal W}_{\text{\sc
eff}}^{\star}$ with the second law $\delta{\mathcal Q}_{\text{\sc
eff}}^{\star} = T_{\text{\sc eff}}^{\star}\,d S_N$ for a reversible
process, we have
\begin{equation}\label{eq:appendix_relation}
    \textstyle T_{\text{\sc eff}}^{\star}\, =\, \left(\frac{\partial U_s}{\partial S_N}\right)_{\partial {\mathcal W}_{\text{\sc eff}}^{\star}=0}\,.
\end{equation}
It is also interesting to note that equation
(\ref{eq:effective_internal_energy1}) can be recovered by the
relation
\begin{equation}
    \textstyle U_s\, =\, U_{\text{\sc eff}}^{\star}\, =\,
    -\frac{\partial}{\beta_{\text{\sc eff}}^{\star}} \ln Z_{\text{\sc eff}}^{\star}\,,
\end{equation}
which is well-defined in terms of a generating function
\begin{equation}\label{eq:appendix_generating_function}
    {\textstyle\frac{\partial}{\partial \beta_{\text{\sc
    eff}}^{\star}} \ln Z_{\text{\sc eff}}^{\star}\, =\, \frac{\partial}{\partial A}} \ln \sum_n \left.e^{A\,\hbar \omega_{\text{\sc
    eff}}^{\star}\,(n + \frac{1}{2})}\right|_{A=\beta_{\text{\sc
    eff}}^{\star}}\,.
\end{equation}
From (\ref{eq:effective_oscillator_entropy1}) it then follows as
well that the effective free energy $F_{\text{\sc eff}}^{\star} =
U_s - T_{\text{\sc eff}}^{\star}\,S_N$, where $F_{\text{\sc
eff}}^{\star} = -k_B T_{\text{\sc eff}}^{\star}\,\ln Z_{\text{\sc
eff}}^{\star}$.

%

%
%
\vspace*{2cm}

\begin{figure}[htb]
\centering\hspace*{-1.5cm}{\includegraphics[scale=0.9]{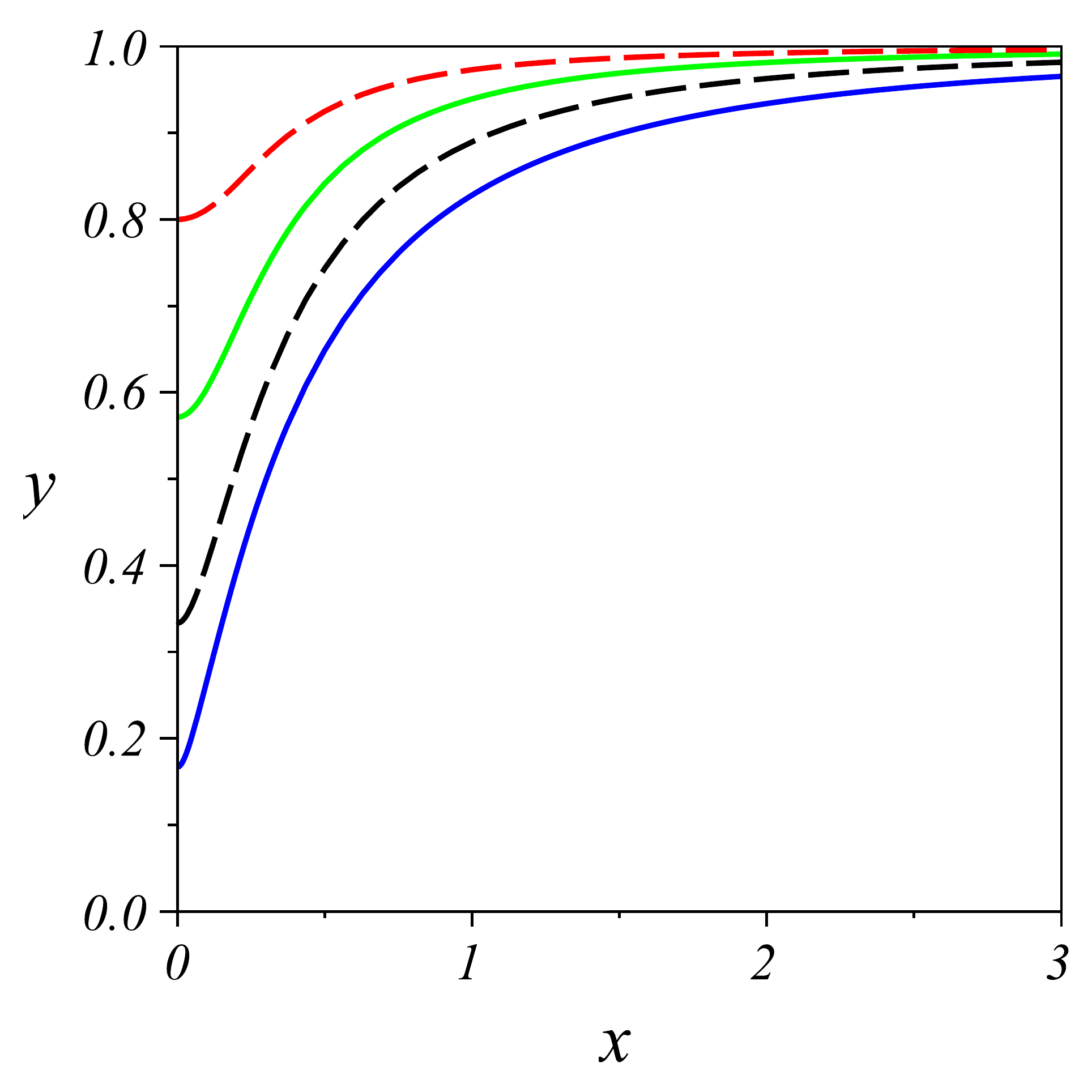}
\caption{\label{fig:fig1}}}
\end{figure}

Fig.~\ref{fig:fig1}: (Color online) $y = k_0/k_{\text{\sc
eff}}^{\star} = 2/(1 + (\Omega +
\gamma)\,\<\hat{p}^2\>_{\beta}/\{\Omega\,(M {\mathbf
w}_0)^2\,\<\hat{q}^2\>_{\beta}\})$ versus $x = k_B T/\hbar {\mathbf
w}_0$ (dimensionless temperature), where ${\mathbf w}_0$ is the
renormalized eigen frequency of the oscillator $\hat{H}_s$ ({\em
cf}. equation (\ref{eq:parameter_change0})); for $k_{\text{\sc
eff}}^{\star}$ refer to equation
(\ref{eq:special_effective_spring_constant1}). From bottom to top,
(blue solid: damping parameter $\gamma = 10$, overdamped), (black
dash: $\gamma = 4$, overdamped), (green solid: $\gamma = 3/2$,
underdamped) and (red dash: $\gamma = 1/2$, underdamped). We have $y
\leq 1$ so that $M/M_{\text{\sc eff}}^{\star} = 1/(2 - y) \leq 1$.
Here $\hbar = k_B = {\mathbf w}_0 = \Omega = M = 1$.\vspace*{.7cm}

\begin{figure}[htb]
\centering\hspace*{-1.5cm}{\includegraphics[scale=0.9]{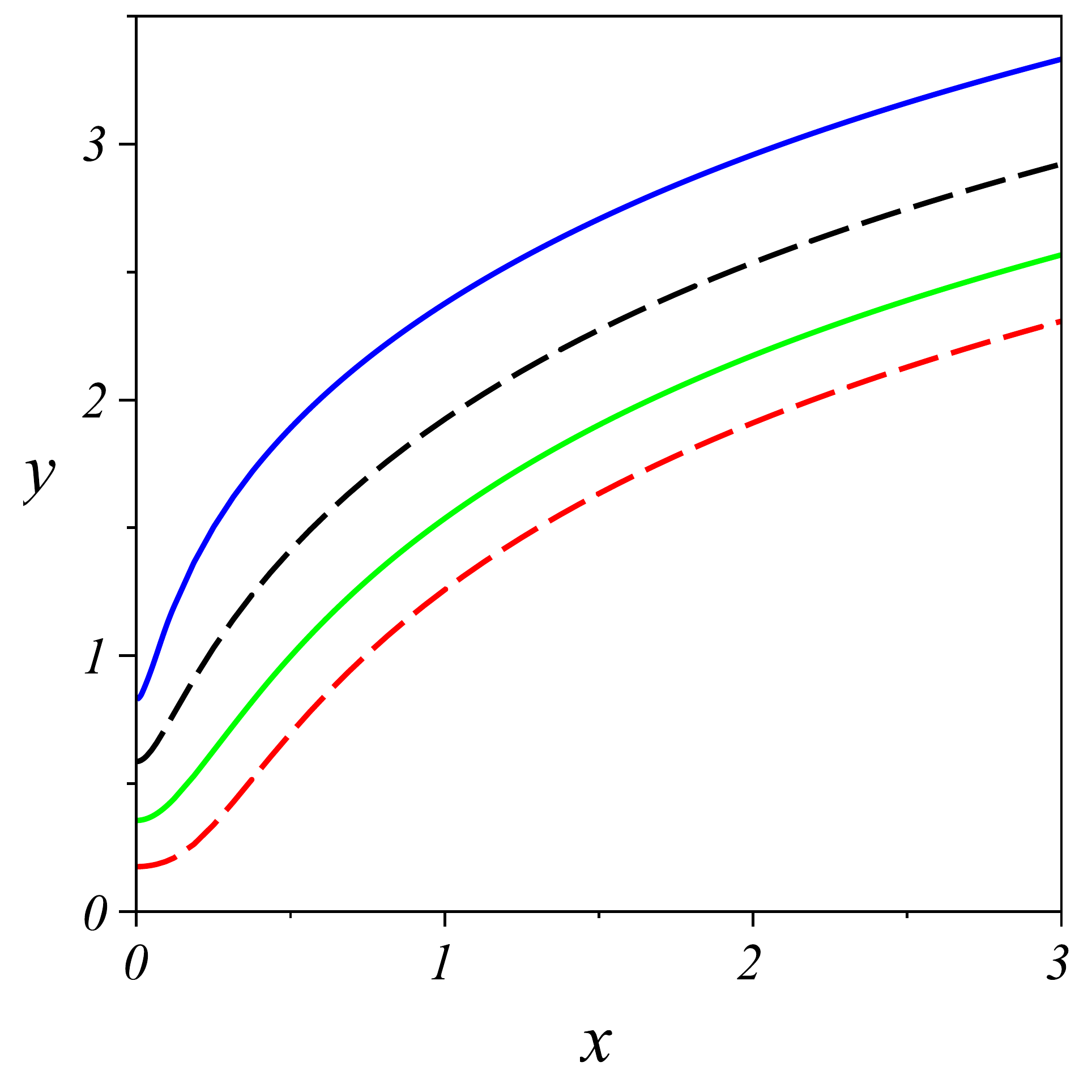}
\caption{\label{fig:fig2}}}
\end{figure}

Fig.~\ref{fig:fig2}: (Color online) $y = S_N$ (von-Neumann entropy)
versus $x = k_B T/\hbar {\mathbf w}_0$ (dimensionless temperature);
for $S_N$ refer to equation (\ref{eq:von-Neumann_entropy1}). From
top to bottom, (blue solid: $\gamma = 10$), (black dash: $\gamma =
4$), (green solid: $\gamma = 3/2$) and (red dash: $\gamma = 1/2$).
As $\gamma$ decreases, then $S_N$ decreases. Here $\hbar = k_B =
{\mathbf w}_0 = \Omega = 1$.\vspace*{.7cm}

\begin{figure}[htb]
\centering\hspace*{-1.5cm}{\includegraphics[scale=0.9]{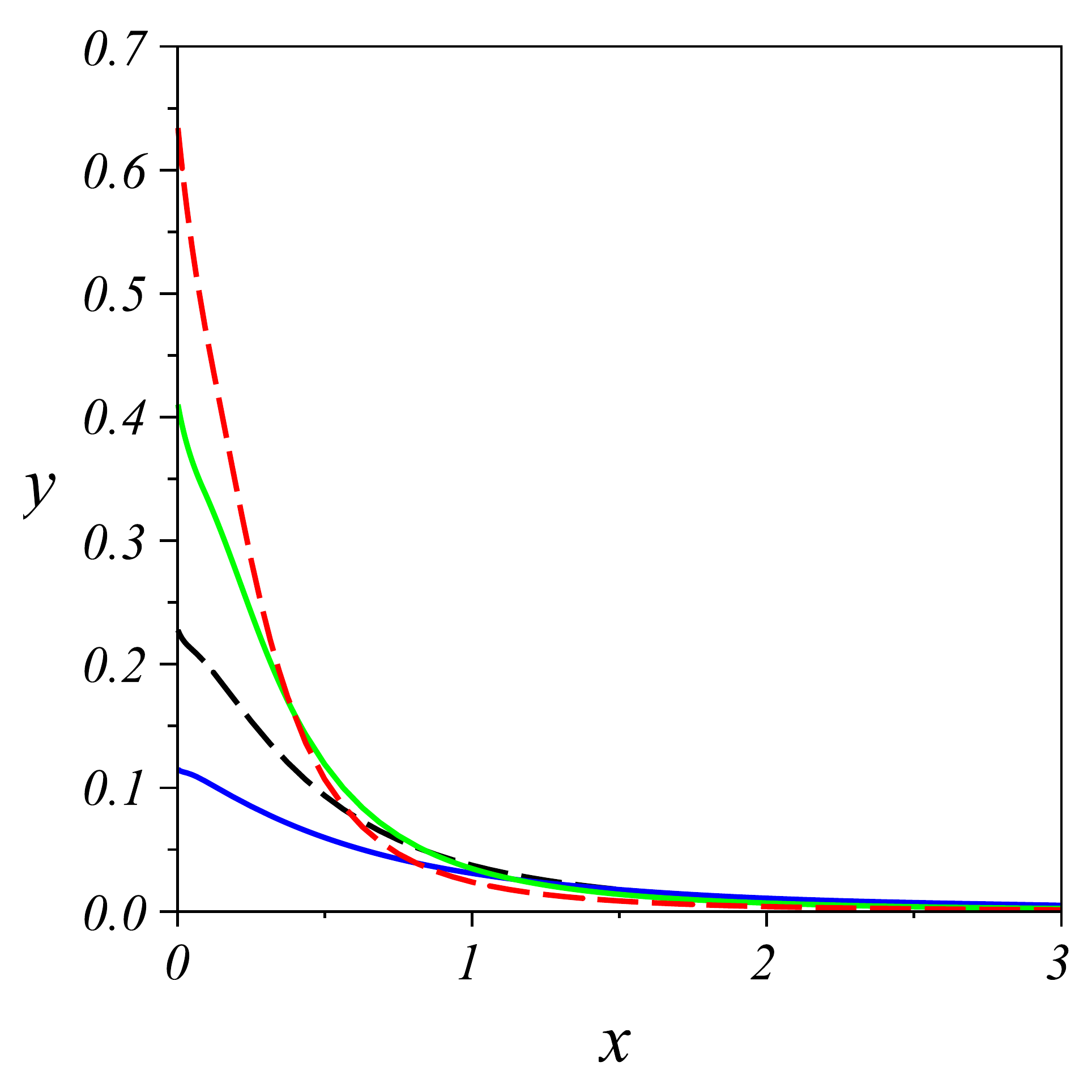}
\caption{\label{fig:fig3}}}
\end{figure}

Fig.~\ref{fig:fig3}: (Color online) $y = 10 \cdot
\left(\partial{\mathcal Q}_{s}/\partial \gamma - T\,\partial
S_N/\partial \gamma\right)/\hbar$ (dimensionless) versus $x = k_B
T/\hbar {\mathbf w}_0$ (dimensionless temperature); for $y$ refer to
equations (\ref{eq:coupling_variation_second_law1}) and
(\ref{eq:coupling_variation_entropy1}). From bottom to top (at $T =
0$), (blue solid: $\gamma = 10$), (black dash: $\gamma = 4$), (green
solid: $\gamma = 3/2$) and (red dash: $\gamma = 1/2$). Here $\hbar =
k_B = {\mathbf w}_0 = \Omega = M = 1$.\vspace*{.7cm}

\begin{figure}[htb]
\centering\hspace*{-1.5cm}{\includegraphics[scale=0.9]{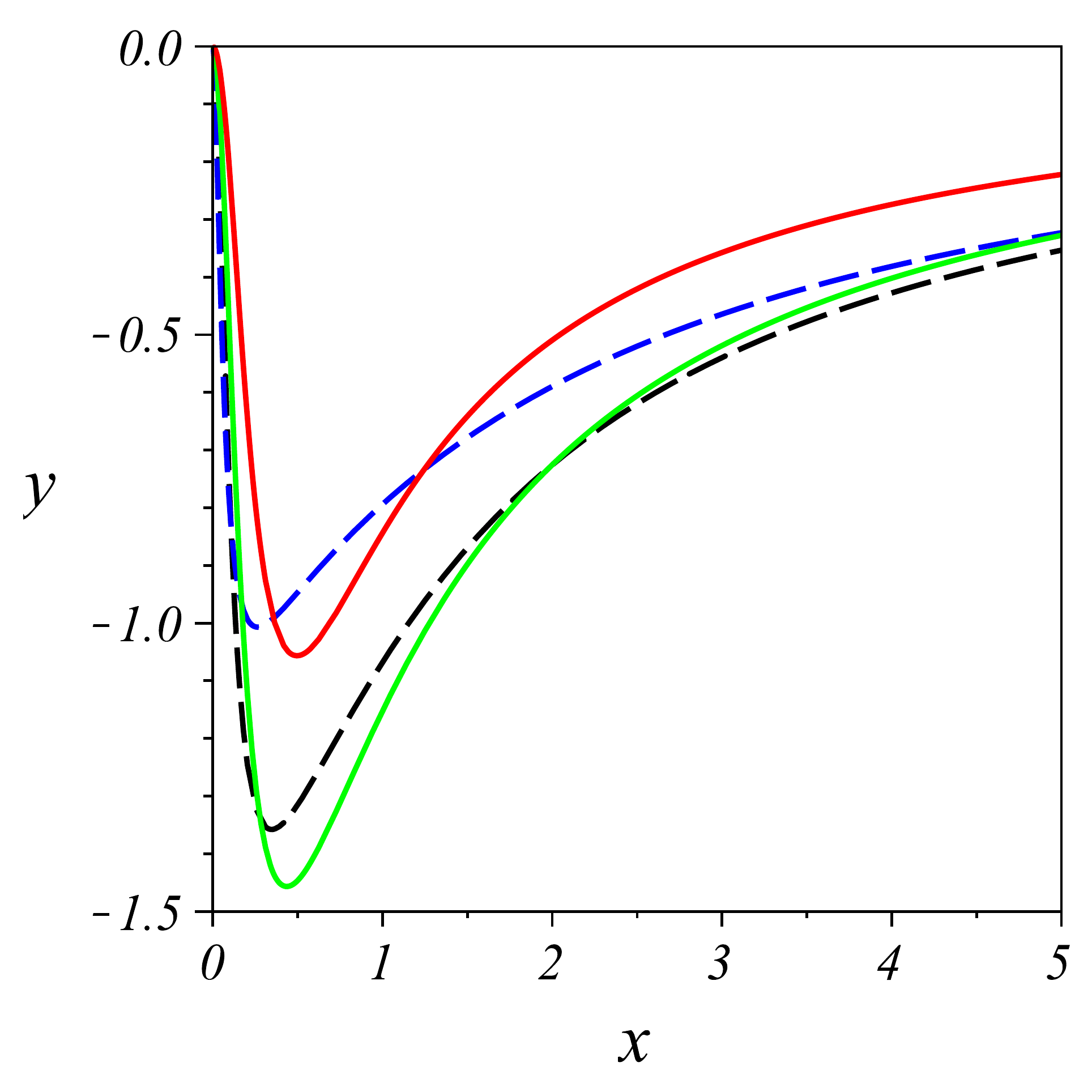}
\caption{\label{fig:fig4}}}
\end{figure}

Fig.~\ref{fig:fig4}: (Color online) $y = 100 \cdot
\left(\partial{\mathcal W}_{\text{\sc eff}}^{\star}/\partial
\gamma\right)/\hbar$ (dimensionless) versus $x = k_B T/\hbar
{\mathbf w}_0$ (dimensionless temperature); for $y$ refer to
equation (\ref{eq:effective_second_law_01}). From top to bottom (at
$T = 0.5$), (blue dash: $\gamma = 10$), (red solid: $\gamma = 1/2$),
(black dash: $\gamma = 4$) and (green solid: $\gamma = 3/2$). Here
$\hbar = k_B = {\mathbf w}_0 = \Omega = M = 1$.\vspace*{.7cm}

\begin{figure}[htb]
\centering\hspace*{-1.5cm}{\includegraphics[scale=0.9]{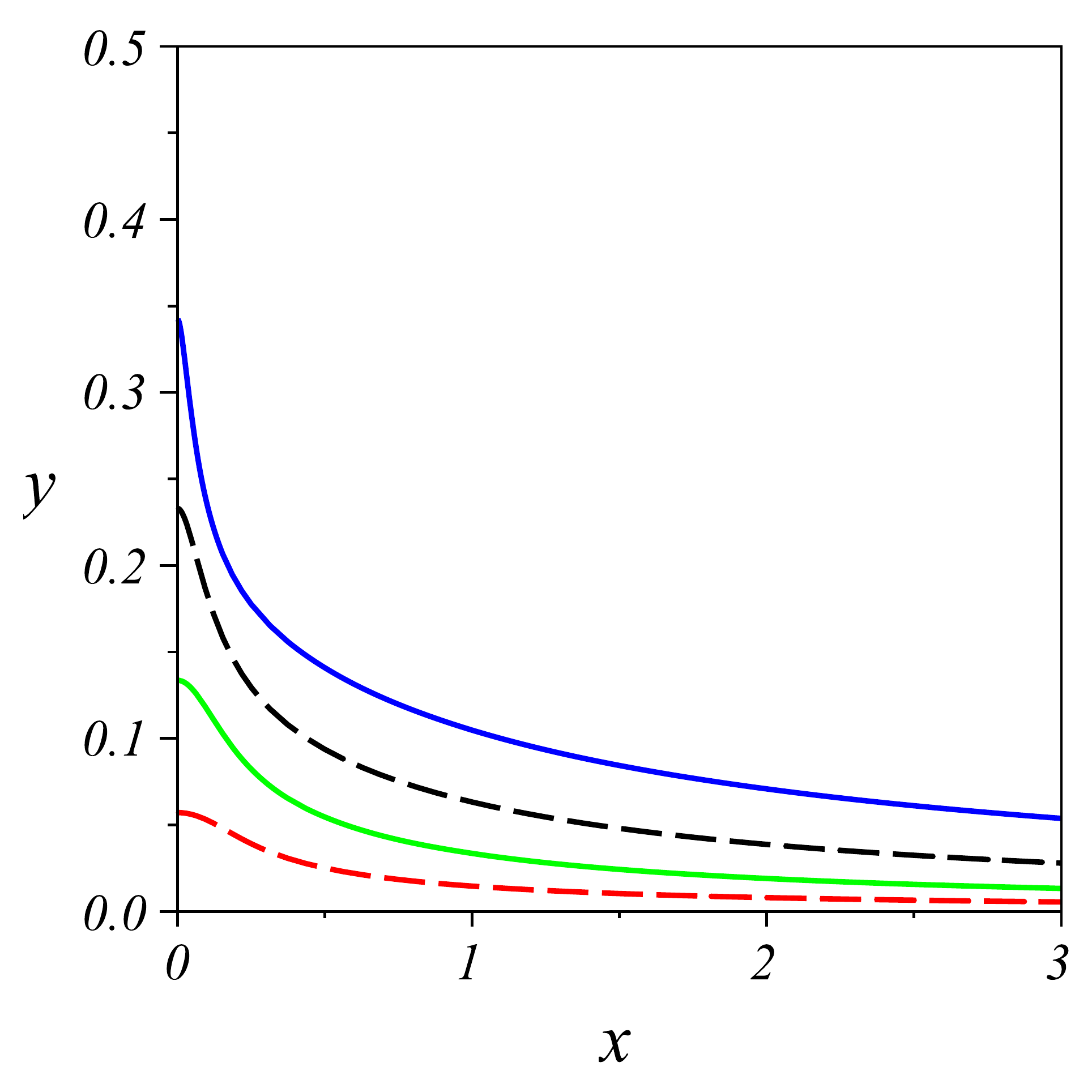}
\caption{\label{fig:fig5}}}
\end{figure}

Fig.~\ref{fig:fig5}: (Color online) $y = \left(\partial{\mathcal
Q}_{s}/\partial M - T_{\text{\sc eff}}^{\star}\,\partial
S_N/\partial M\right)/(\hbar {\mathbf w}_0/M)$ (dimensionless)
versus $x = k_B T/\hbar {\mathbf w}_0$ (dimensionless temperature);
for $y$ refer to equation (\ref{eq:second_law_well_defined1}). From
top to bottom, (blue solid: $\gamma = 10$), (black dash: $\gamma =
4$), (green solid: $\gamma = 3/2$) and (red dash: $\gamma = 1/2$).
As $\gamma$ decreases, then $y$ decreases. Here $\hbar = k_B =
{\mathbf w}_0 = \Omega = M = 1$.\vspace*{.7cm}

\begin{figure}[htb]
\centering\hspace*{-1.5cm}{\includegraphics[scale=0.9]{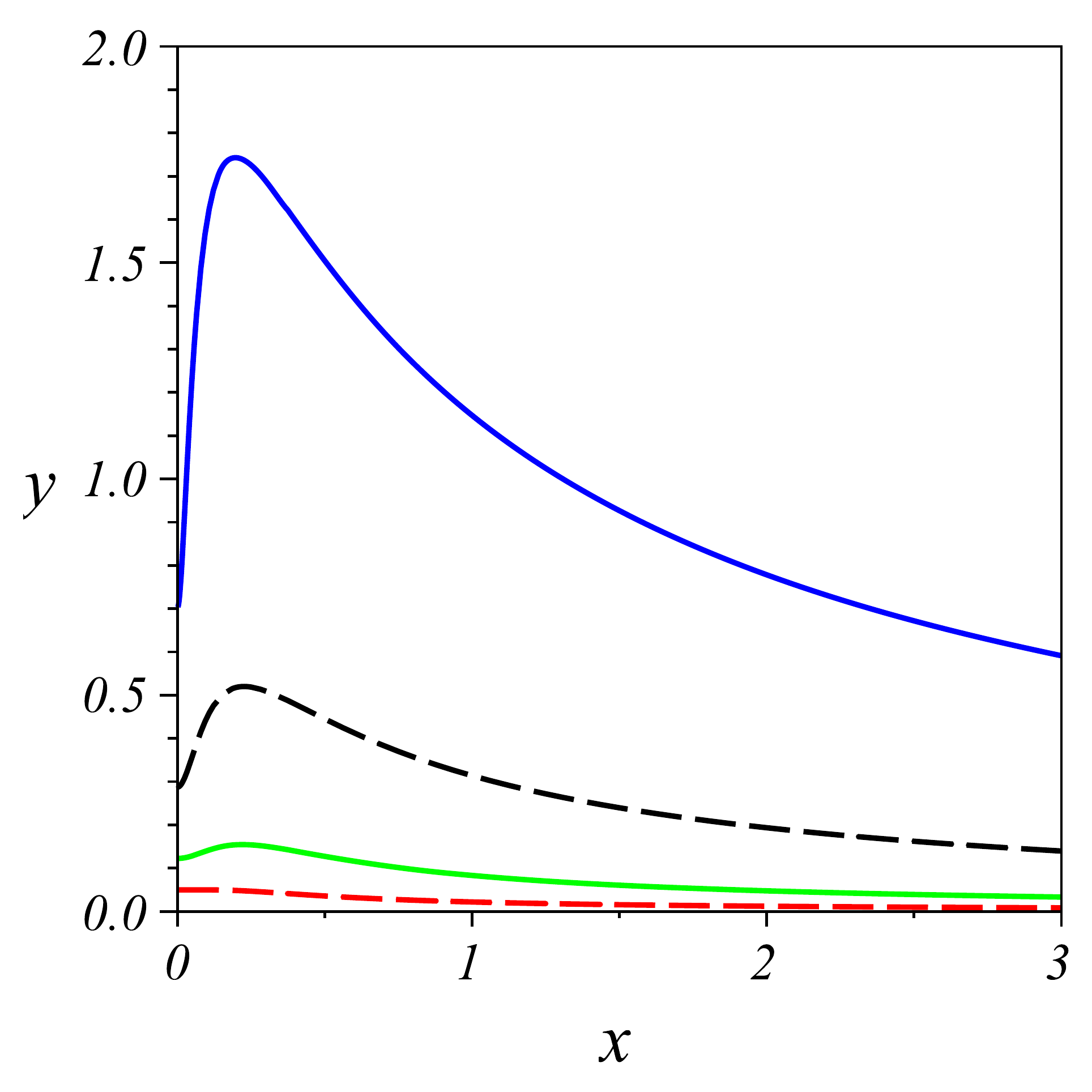}
\caption{\label{fig:fig6}}}
\end{figure}

Fig.~\ref{fig:fig6}: (Color online) $y = \left(\partial{\mathcal
Q}_{s}/\partial k_0 - T_{\text{\sc eff}}^{\star}\,\partial
S_N/\partial k_0\right)/(\hbar/(M {\mathbf w}_0))$ (dimensionless)
versus $x = k_B T/\hbar {\mathbf w}_0$ (dimensionless temperature);
for $y$ refer to equation (\ref{eq:second_law_well_defined2}). From
top to bottom, (blue solid: $\gamma = 10$), (black dash: $\gamma =
4$), (green solid: $\gamma = 3/2$) and (red dash: $\gamma = 1/2$).
As $\gamma$ decreases, then $y$ decreases. Here $\hbar = k_B =
{\mathbf w}_0 = \Omega = M = 1$.\vspace*{.7cm}

\begin{figure}[htb]
\centering\hspace*{-1.5cm}{\includegraphics[scale=0.9]{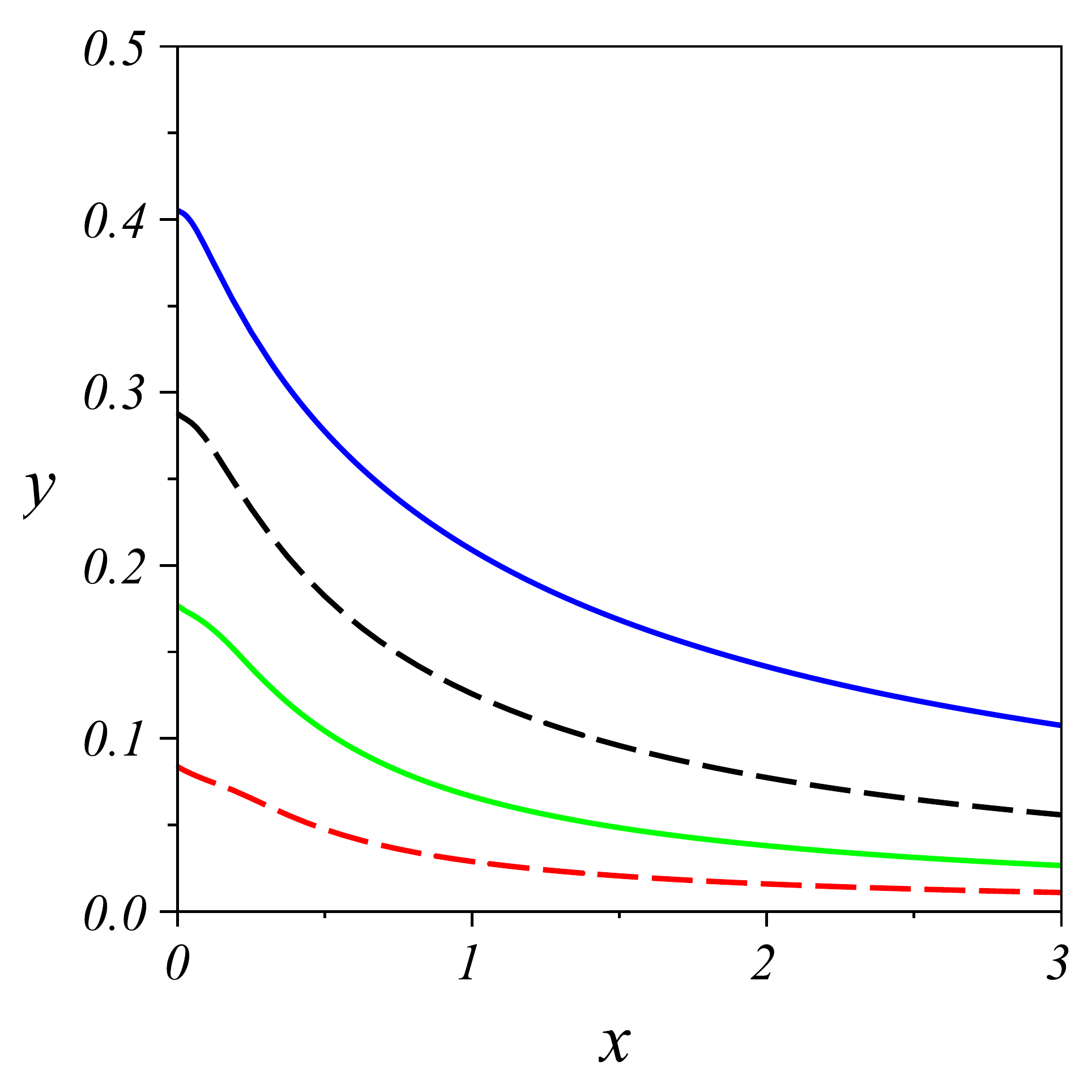}
\caption{\label{fig:fig7}}}
\end{figure}

Fig.~\ref{fig:fig7}: (Color online) $y = \left(\partial{\mathcal
Q}_{s}/\partial M - T\,\partial S_N/\partial M\right)/(\hbar
{\mathbf w}_0/M)$ (dimensionless) versus $x = k_B T/\hbar {\mathbf
w}_0$ (dimensionless temperature); for $y$ refer to equation
(\ref{eq:second_law2_21}). From top to bottom, (blue solid: $\gamma
= 10$), (black dash: $\gamma = 4$), (green solid: $\gamma = 3/2$)
and (red dash: $\gamma = 1/2$). As $\gamma$ decreases, then $y$
decreases. Here $\hbar = k_B = {\mathbf w}_0 = \Omega = M = 1$.


\begin{thebibliography}{1}
%
\bibitem{SHE02} D.P. Sheehan,
{\em Quantum limits to the second law}, AIP Conference Proceedings
{\bf 643} (2002).
%
\bibitem{MAH04} J. Gemmer, M. Michel and G. Mahler, {\em Quantum
Thermodynamics} (Springer, Berlin, 2004).
%
\bibitem{CAP05} C. Vladislav and D.P. Sheehan,
{\em Challenges to the second law of thermodynamics: theory and
experiment} (Springer, New York, 2005).
%
\bibitem{ALL00} A.E. Allahverdyan and Th.M. Nieuwenhuizen, Phys.
Rev. Lett. {\bf 85}, 1799 (2000).
%
\bibitem{NIE02} Th.M. Nieuwenhuizen and A.E. Allahverdyan, Phys.
Rev. E {\bf 66}, 036102 (2002).
%
\bibitem{ANN02} K. Annamalai and I.K. Puri, {\em Advanced
Thermodynamics Engineering} (CRC Press, New York, 2002).
%
\bibitem{FOR06} G.W. Ford and R.F. O'Connell, Phys. Rev. Lett. {\bf 96}, 020402 (2006).
%
\bibitem{KIM06} I. Kim and G. Mahler, Eur. Phys. J. B {\bf 54}, 405 (2006).
%
\bibitem{KIM07} I. Kim and G. Mahler, Eur. Phys. J. B {\bf 60}, 401 (2007).
%
\bibitem{BUE04} A.N. Jordan and M. B\"{u}ttiker, Phys. Rev. Lett.
{\bf 92}, 247901 (2004).
%
\bibitem{BUE05} M. B\"{u}ttiker and A.N. Jordan, Physica E {\bf 29},
272 (2005).
%
\bibitem{LUT08} S. Hilt and E. Lutz, Phys. Rev. A {\bf 79}, 010101(R) (2009). Here the Rubin model of
the total Hamiltonian was considered while in the current paper the
model Hamiltonian (\ref{eq:total_hamiltonian1}) with
(\ref{eq:total_hamiltonian2}) is under consideration.
%
\bibitem{THU07} R. Hanel and S. Thurner, Physica A {\bf 380}, 109 (2007).
%
\bibitem{ABE08} S. Abe and S. Thurner, Europhys. Lett. {\bf 81}, 10004 (2008).
%
\bibitem{CAL51} H.B. Callen and T.A. Welton, Phys. Rev. {\bf 83}, 34 (1951).
%
\bibitem{WEI99} U. Weiss, {\em Quantum dissipative systems}, 2nd edn.
(World Scientific, Singapore, 1999).
%
\bibitem{FOR85} G.W. Ford, J.T. Lewis and R.F. O'Connell, Phys. Rev.
Lett. {\bf 55}, 2273 (1985).
%
\bibitem{ABS74} M. Abramowitz and I. Stegun, {\em Handbook of Mathematical Functions
with Formulas, Graphs, and Mathematical Tables} (Dover, New York,
1974).
%
\bibitem{ING88} H. Grabert, P. Schramm and G.-L. Ingold, Phys. Rep.
{\bf 168}, 115 (1988).
%
\bibitem{FEY98} R.P. Feynman, {\em Statistical Mechanics: A Set of Lectures}
(Addison-Wesley, Reading, 1998).
%
\bibitem{BUE08} C. H\"{o}rhammer and H. B\"{u}ttner, J. Stat. Phys.
{\bf 133}, 1161 (2008).
%
\bibitem{SRE93} M. Srednicki, Phys. Rev. Lett. {\bf 71}, 666 (1993).
%
\bibitem{GRA07} I.S. Gradshteyn and I.M. Ryzhik, {\em Table of Integrals, Series, and Products}, 7th edn. (Academic
Press, San Diego, 2007).
%
\bibitem{GRA84} H. Grabert, U. Weiss, P. Talkner, Z. Phys. B {\bf
55}, 87 (1984).
%
\bibitem{HAE05} P. H\"anggi and G.-L. Ingold, Chaos, {\bf 15}, 026105 (2005).
%
\bibitem{FOR05} G.W. Ford and R.F. O'Connell, Physica E {\bf 29}, 82 (2005).
%
\bibitem{HAE06} P. H\"anggi and G.-L. Ingold, Acta Physica Polonica B {\bf 37}, 1537 (2006).
%
\bibitem{JAY57} E. Jaynes, Phys Rev {\bf 106}, 106 (1957).
%
\bibitem{EJA57} E. Jaynes, Phys. Rev {\bf 108}, 171 (1957).
%
\bibitem{LAN61} R. Landauer, IBM J. Res. Dev. {\bf 5}, 183 (1961).
%
\bibitem{BEN03} C.H. Bennett, Stud. Hist. Philos. Mod. Phys. {\bf 34}, 501
(2003), which is also available as arXiv:physics/0210005.
%
\bibitem{KIM09} I. Kim and G. Mahler, in preparation.
%
\bibitem{SZE39} G. Szeg\"{o}, {\em Orthogonal polynomials}
(American Mathematical Society, New York, 1939).
%
\end{thebibliography}
\end{document}